\documentclass[preprint]{aastex}
\usepackage{graphicx}     
\usepackage{natbib}
\usepackage{amsmath,amsthm,amssymb}
\usepackage{color}
\usepackage{ulem}
\newcommand{\gps}{\ensuremath{g_{\rm P1}}}
\newcommand{\rps}{\ensuremath{r_{\rm P1}}}
\newcommand{\ips}{\ensuremath{i_{\rm P1}}}
\newcommand{\zps}{\ensuremath{z_{\rm P1}}}
\newcommand{\yps}{\ensuremath{y_{\rm P1}}}

\newcommand{\PS}{\protect \hbox {Pan-STARRS1}}
\newcommand{\WISE}{{\it WISE}}
\newcommand{\iz}{\ips$-$\zps}
\newcommand{\iy}{\ips$-$\yps}
\newcommand{\zy}{\zps$-$\yps}
\newcommand{\ywa}{\yps$-W1$}
\newcommand{\wawb}{$W1-W2$}
\newcommand{\wbwc}{$W2-W3$}
\newcommand{\mytilde}{\raise.17ex\hbox{$\scriptstyle\mathtt{\sim}$}}

\shorttitle{Nearby L/T Dwarfs Including Variables}
\shortauthors{Best, W. M. J. et al}

\slugcomment{\large Accepted by {\it The Astrophysical Journal}, 2013 August 29}

\begin{document}

\title{A Search for L/T Transition Dwarfs With \PS\ and \WISE: \\
  Discovery of Seven Nearby Objects Including Two Candidate Spectroscopic Variables}
\author{William M. J. Best\altaffilmark{1}, Michael C. Liu\altaffilmark{1,6},
  Eugene A. Magnier\altaffilmark{1}, Kimberly M. Aller\altaffilmark{1,6}, Niall
  R. Deacon\altaffilmark{2}, Trent J. Dupuy\altaffilmark{3}, Joshua Redstone\altaffilmark{4},
W. S. Burgett\altaffilmark{1}, 
K. C. Chambers\altaffilmark{1},
K. W. Hodapp\altaffilmark{1}, 
N. Kaiser\altaffilmark{1}, 
R.-P. Kudritzki\altaffilmark{1},
J. S. Morgan\altaffilmark{1},
P. A. Price\altaffilmark{5},
J. L. Tonry\altaffilmark{1}, and
R. J. Wainscoat\altaffilmark{1}}
\altaffiltext{1}{Institute for Astronomy, University of Hawaii at Manoa,
  Honolulu, HI 96822, USA; wbest@ifa.hawaii.edu}
\altaffiltext{2}{Max Planck Institute for Astronomy, Koenigstuhl 17, D-69117 Heidelberg, Germany}
\altaffiltext{3}{Harvard-Smithsonian Center for Astrophysics, 60 Garden Street, Cambridge, MA 02138, USA}
\altaffiltext{4}{Facebook, 335 Madison Ave, New York, NY 10017-4677, USA}
\altaffiltext{5}{Department of Astrophysical Sciences, Princeton University, Princeton, NJ 08544, USA}
\altaffiltext{6}{Visiting Astronomer at the Infrared Telescope Facility, which
  is operated by the University of Hawaii under Cooperative Agreement
  no. NNX-08AE38A with the National Aeronautics and Space Administration,
  Science Mission Directorate, Planetary Astronomy Program.}

\begin{abstract}
  We present initial results from a wide-field (30,000~deg$^2$) search for L/T
  transition brown dwarfs within 25~pc using the \PS\ and \WISE\ surveys.
  Previous large-area searches have been incomplete for L/T transition dwarfs,
  because these objects are faint in optical bands and have near-infrared colors
  that are difficult to distinguish from background stars.  To overcome these
  obstacles, we have cross-matched the \PS\ (optical) and \WISE\ (mid-IR)
  catalogs to produce a unique multi-wavelength database for finding ultracool
  dwarfs.  As part of our initial discoveries, we have identified seven brown
  dwarfs in the L/T transition within $9-15$~pc of the Sun.  The L9.5 dwarf
  PSO~J140.2308+45.6487 and the T1.5 dwarf PSO~J307.6784+07.8263 \citep[both
  independently discovered by][]{Mace:2013jh} show possible spectroscopic
  variability at the $Y$- and $J$-bands.  Two more objects in our sample show
  evidence of photometric $J$-band variability, and two others are candidate
  unresolved binaries based on their spectra.  We expect our full search to
  yield a well-defined, volume-limited sample of L/T transition dwarfs that will
  include many new targets for study of this complex regime.
  PSO~J307.6784+07.8263 in particular may be an excellent candidate for in-depth
  study of variability, given its brightness ($J=14.2$~mag) and proximity
  (11~pc).
\end{abstract}


\section{Introduction}
\label{intro}
Over 1,200 brown dwarfs have been cataloged since the first unambiguous
discovery less than twenty years ago \citep{Nakajima:1995bb}.  As a result, two
new spectral types, L and T, have been created
\citep{Kirkpatrick:1999ev,Burgasser:2006cf} to categorize these ultracool
objects, and recent discoveries have identified the first brown dwarfs cooler
than 400 K \citep{Liu:2011hb,Luhman:2011jo,Cushing:2011dk,Kirkpatrick:2012ha},
spawning the creation of the Y spectral type.  The main drivers of brown dwarf
discoveries have been large-area digital surveys such as the Deep Near Infrared
Survey of the Southern Sky \citep[DENIS,][]{Epchtein:1999tz}, the Sloan Digital
Sky Survey \citep[SDSS;][]{York:2000gn}, the Two Micron All Sky Survey
\citep[2MASS;][]{Skrutskie:2006hl}, and the UKIRT Infrared Deep Sky Survey
\citep[UKIDSS;][]{Lawrence:2007hu}.  Photometric searches using these surveys
combined with spectroscopic follow-up have identified most of the known field
brown dwarfs \citep[e.g.,][]{Chiu:2006jd,Cruz:2007kb,Burningham:2010dh}.  More
recently, the \PS\ 3$\pi$ Survey \citep[PS1;][]{Kaiser:2010gr}, the Wide-Field
Infrared Survey Explorer \citep[\WISE;][]{Wright:2010in}, and the VISTA
Hemisphere Survey (VHS; PI McMahon, Cambridge, UK) have pushed brown dwarf
searches to greater distances and cooler temperatures
\citep[e.g.,][]{Deacon:2011gz,Kirkpatrick:2011ey,Lodieu:2012go}.

Despite the overall success of past searches, it is expected that previous work
has been less complete for L/T transition dwarfs (spectral types
$\approx$\,L6--T5) because these objects are optically faint and have
near-infrared colors that are difficult to distinguish from M and early-L dwarfs
\citep[e.g.,][]{Reid:2008fz}.  Past searches sensitive to L/T objects have also
focused on modest areas of the sky.  The most successful searches so far have
been those of \citet{Chiu:2006jd}, who searched over 3,500~deg$^2$ using SDSS
photometry (optical $iz$ bands) to find 47 L6--T5 dwarfs; \citet{Metchev:2008gx}
and \citet{Geissler:2011gg}, who cross-matched SDSS DR1 \citep[2,099
deg$^2$;][]{Abazajian:2003cn} with 2MASS (near-IR) photometry to identify 10
L6--T5 dwarfs (many of which showed spectral indications of binarity); and
\citet{DayJones:2013hm}, who searched 495~deg$^2$ in UKIDSS and SDSS to find 15
L6--T5 dwarfs.  \citet[][hereinafter D11]{Deacon:2011gz} and
\citeauthor{Liu:2011hc} \!\!(2011a; 2013, in prep) have combined \PS\ and 2MASS
photometry with proper motion to search three-quarters of the sky for T dwarfs.
To date, however, there has been no large-area search specifically targeting
nearby, bright L/T transition dwarfs.

Because of this slow progress in identifying the local L/T population, we do not
yet have a well-constrained substellar mass function for the solar neighborhood.
However, recent work indicates that the local space density has a minimum at the
L/T transition (\citealp{DayJones:2013hm} and references therein).  A natural
explanation is that the L/T transition spans a fairly narrow temperature range
($\Delta T\approx400$~K) compared to L0--L5 dwarfs \citep[$\Delta
T\approx1000$~K;][]{Golimowski:2004en,Stephens:2009cc}.  Moreover, brown dwarfs
cool as they age and evolve to later spectral types, resulting in a growing
accumulation of late-T and Y dwarfs \citep{Burgasser:2004ed}.
\citet{DayJones:2013hm} calculate errors of $\sim\!17\%$ and $\sim\!40\%$ for
their estimates of the space density of L7--T0.5 and T1--T4.5 dwarfs,
respectively.  A larger sample of L/T objects will improve these estimates,
allowing more accurate assessments of the local ultracool IMF.

L/T transition dwarfs are of further interest because their spectral features
indicate significant variations in surface gravity, metallicity, and/or
atmospheric clouds (\citealp{Kirkpatrick:2005cv} and references therein).  In
particular, many changes in these spectral features are thought to arise from
the depletion of condensate clouds as brown dwarfs cool
\citep[e.g.,][]{Allard:2001fh,Burrows:2006ia,Saumon:2008im}, making L/T
transition dwarfs ideal for case studies of cloud formation and gas chemistry in
ultracool atmospheres.  One notable example is the $\sim\!0.5$~mag $J$-band
brightening across the L/T transition as thick clouds dissipate to reveal lower,
warmer layers of the photosphere, observed in isolated objects
\citep{Tinney:2003eg,Dupuy:2012bp} as well as the components of L/T binaries
\citep{Liu:2006ce,Looper:2008if}.

Another likely consequence of evolving and thinning clouds is photometric
variability.  Optical variability is known to be common in L dwarfs
\citep{Rockenfeller:2006ij} and is typically periodic on timescales of several
hours, consistent with rotation periods of brown dwarfs
\citep{BailerJones:2004ks,Reiners:2008kf}.  However, detection of infrared
variability has proven to be much more elusive
(\citealp[e.g.,][]{Enoch:2003ce,Koen:2005cx,Goldman:2008es}).  The first
unambiguous detections of near-IR photometric variability in T dwarfs were in
SIMP J013656.57+093347.3 \citep[T2.5;][hereinafter
SIMP~0136+0933]{Artigau:2009bk} and 2MASS~J21392676+0220226
\citep[T1.5;][hereinafter 2MASS~2139+0220]{Radigan:2012ki}.  Notably, both of
these objects are members of the L/T transition.  \citet{Apai:2013fn} have
subsequently detected $J$- and $H$-band spectral variability in these objects.
\citet{Buenzli:2012gd} discovered multiband infrared sinusoidal variability in
the T6.5 dwarf 2MASS 2228$-$4310, along with a wavelength dependence of the
variability phase, indicative of vertical cloud structure.  Most recently,
\citet{Khandrika:2013fs} identified photometric $J$-band variability in two
mid-L dwarfs and $K$-band variability in a T8 dwarf, while also confirming the
$J$-band variability of 2MASS~2139+0220.  These observations collectively
suggest that variability is a normal feature of ultracool dwarfs.
\citet{Khandrika:2013fs} estimate a variability fraction of $35\%\pm5\%$, and
surprisingly find no strong evidence of a greater frequency for L/T transition
objects.

Any search for brown dwarfs, especially those in the L/T transition, must also
account for binaries.  L+T binaries can have colors and composite spectra that
mimic those of single L/T transition dwarfs \citep{Burgasser:2007fl} and can
exaggerate the amplitude of the $J$-band brightening
\citep{Liu:2006ce,Dupuy:2012bp}.  Components of binary systems are valuable
benchmarks, as they are coeval, equidistant, and have common metallicities.
They are especially valuable in the L/T transition, where they can help to set
tight constraints on the observed atmospheric transitions
\citep{Dupuy:2009ga,Liu:2010cw}.  Compared to their stellar analogs, field brown
dwarf binaries tend to have smaller separations \citep{Allen:2007hd} and mass
ratios closer to unity \citep{Allen:2007cn}.  This enables determination of
their orbits and dynamical masses
\citep[e.g.,][]{Bouy:2004jk,Liu:2008ib,Dupuy:2010ch,Konopacky:2010kr}, which
breaks the mass/age degeneracy that plagues field brown dwarf analysis.

To increase the census of L/T transition dwarfs, we have begun a search using
PS1 and \WISE, leveraging the combined data of these optical and mid-infrared
surveys.  In order to construct a well-defined sample of L/T transition dwarfs,
we have limited our search to candidates within 25 parsecs of the Sun, based on
their photometric distances.  Several past projects, including the Gliese
catalog \citep{Gliese:1995wz} and the PMSU M dwarf survey \citep[and references
therein]{Reid:2002eh}, have searched this same volume of space for stars, so our
project is well-matched to previous efforts.  Our search will address a known
deficiency in the solar neighborhood census, significantly improve the
constraints on the local substellar mass and luminosity functions, and identify
a well-defined, volume-limited sample of late L and early T dwarfs that can be
used to better understand the evolution of brown dwarfs through the L/T
transition.

In this paper, we present the first results of our ongoing search, namely the
discovery of seven L/T transition dwarfs within 15~pc.  Three objects are
entirely new discoveries.  The other four were identified independently by
\citet[][hereinafter M13]{Mace:2013jh} in their search for T dwarfs in \WISE,
though without distance estimates.  (Our spectroscopic confirmation preceded the
publication of M13.)  We explain our search process in Section~\ref{method}.  We
describe our observations in Section~\ref{obser}, and our discoveries in
Section~\ref{results}.  We discuss the properties of specific discoveries in
Section~\ref{discuss}, and summarize our findings in Section~\ref{summary}.

\section{Search Method}
\label{method} We identified candidate L/T dwarfs through a series of quality,
color, and magnitude cuts applied to our merged PS1+\WISE\ database, followed by
visual inspection of PS1, 2MASS, and \WISE\ images.  We then obtained and typed
spectra of candidates using standard procedures described in
Section~\ref{obser}.

The PS1 $3\pi$ survey (Chambers et al., in prep) is obtaining multi-epoch
imaging in five optical bands ($g_{\rm P1}, r_{\rm P1}, i_{\rm P1}, z_{\rm P1},
y_{\rm P1}$) with a 1.8-meter wide-field telescope on Haleakala, Maui, covering
the entire sky north of declination~$-30^{\circ}$.  Images are processed nightly
through the Image Processing Pipeline
\citep[IPP;][]{Magnier:2006uj,Magnier:2007wn,Magnier:2008jf}, with photometry on
the AB magnitude scale \citep{Tonry:2012gq}.  Imaging began in May 2010 and
should be completed by early 2014.  The \WISE\ All-Sky Release includes data
taken between January and August 2010 \citep{Cutri:2012wz} in four mid-infrared
bands: $W1$ ($3.6\,\mu$m), $W2$ ($4.5\,\mu$m), $W3$ ($12\,\mu$m), and $W4$
($22\,\mu$m).  We merged all PS1 detections through January 2012 with the \WISE\
All-Sky catalog using a $3.0''$ matching radius.  Because the two surveys are
nearly contemporaneous, matching by position is effective for all but the
highest proper motion objects.  Matching the two surveys by position also
eliminates transient objects (e.g., asteroids) in regions the surveys have only
covered once.  Candidates were selected from the full three-quarters of sky
covered by the merged database.

\begin{figure}
\begin{center}
  \includegraphics[width=1.00\columnwidth, trim = 20mm 0 5mm 0]{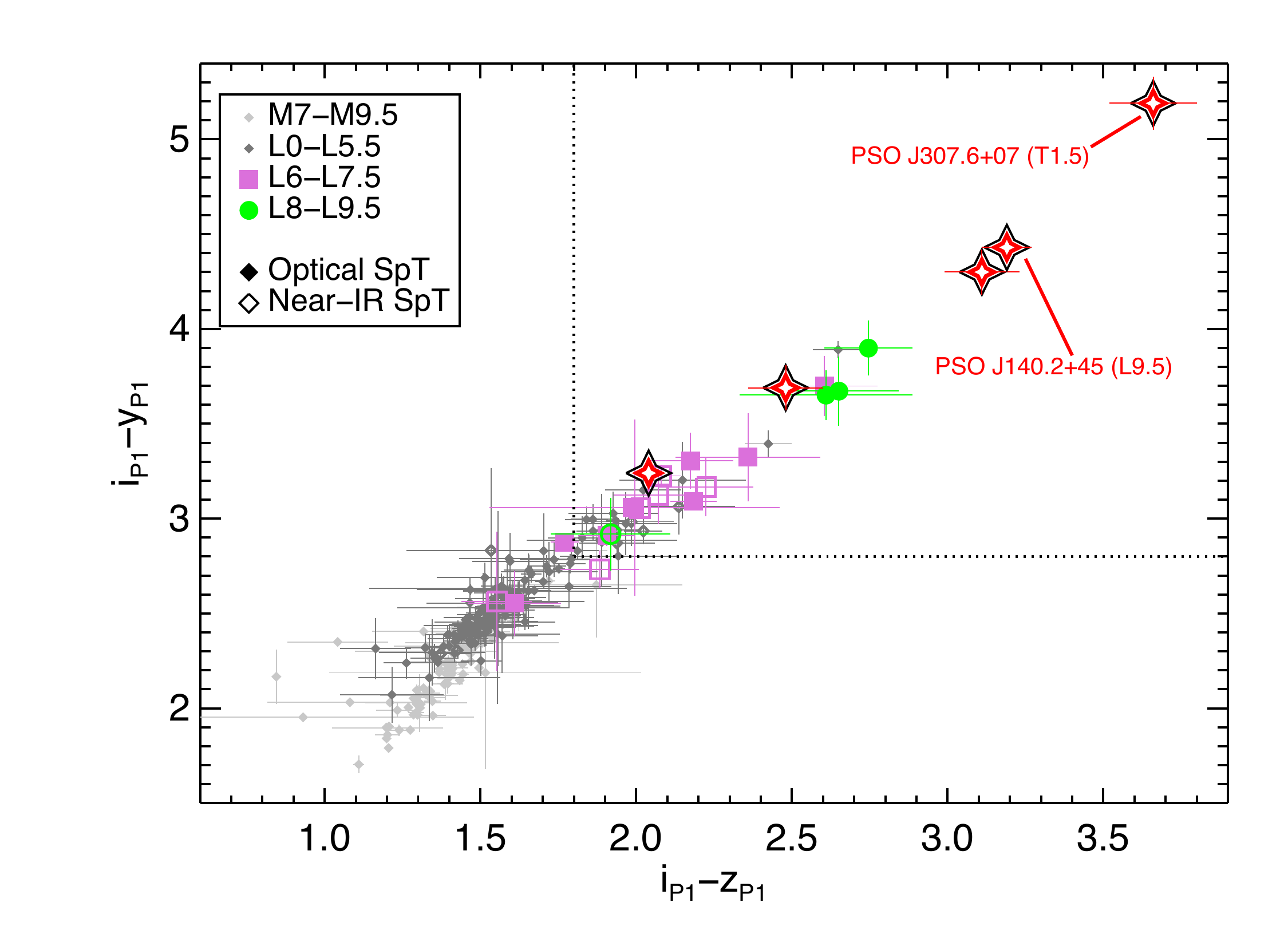}
  \caption{\iz\ vs. \iy\ diagram for known ultracool dwarfs.  Late-L dwarfs are
    indicated with magenta squares (types L6--L7.5) and green circles
    (L8--L9.5); no known T dwarfs have \ips\ detections in our PS1+\WISE\
    database. Objects with optical spectral types are plotted with filled
    symbols, and objects with near-infrared spectral types are plotted with open
    symbols.  The dotted black lines indicate the color cuts used in our search;
    we selected objects above and to the right of the dotted lines, but only
    enforced each cut for objects which had $\sigma<0.2$~mag and at least two
    detections in both bands.  Our newly identified objects are marked with
    large red four-point stars, with the spectroscopic variable candidates
    labeled individually.  (Two of our discoveries and many known L dwarfs were
    also not detected in \ips.)  The new discoveries clearly extend the linear
    \iz\ vs. \iy\ locus to redder colors.}
\label{fig.iy.iz}
\end{center}
\end{figure}

\begin{figure}
\begin{center}
  \includegraphics[width=1.00\columnwidth, trim = 20mm 0 5mm 0]{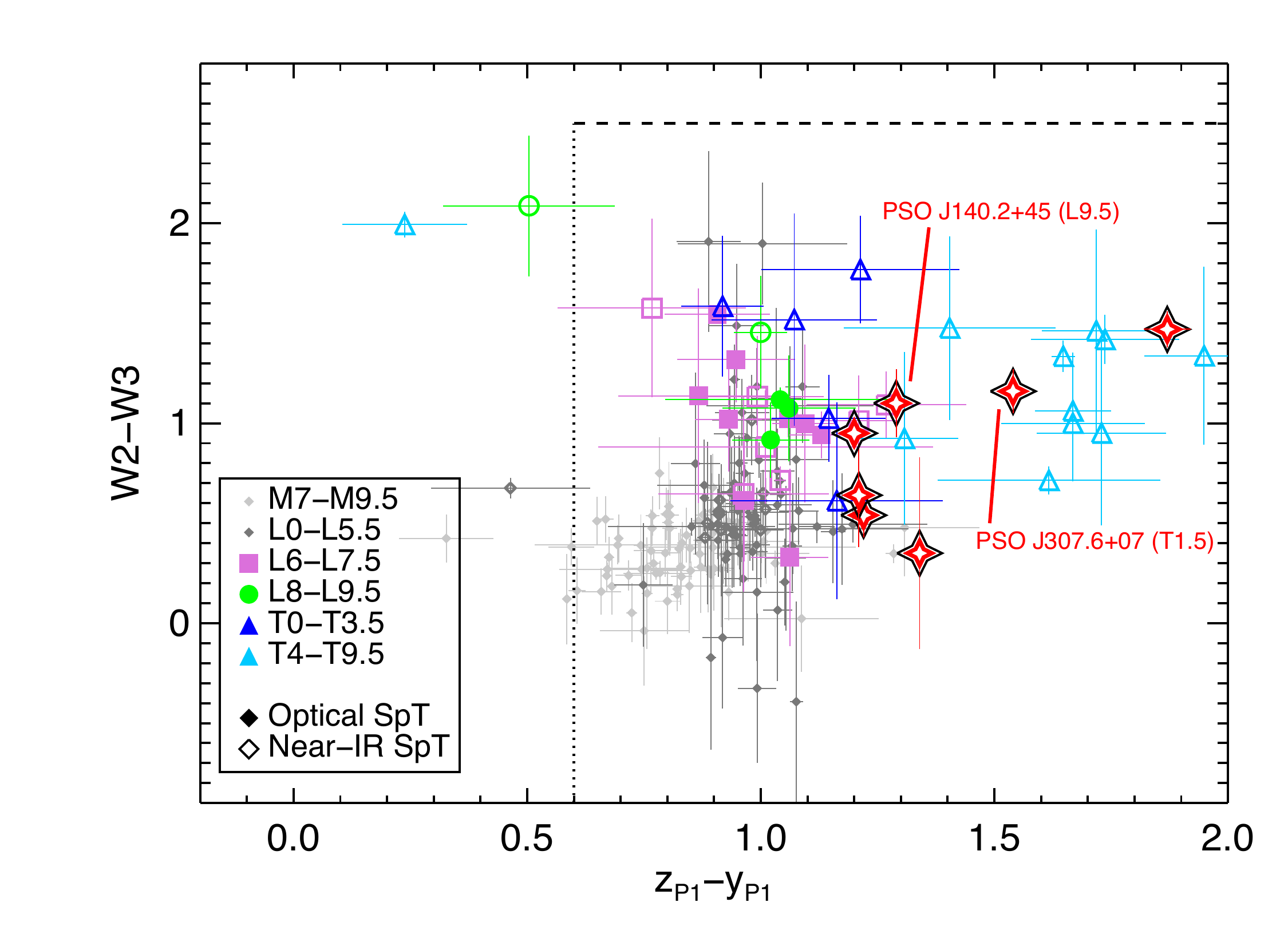}
  \caption{ \wbwc\ vs. \zy\ diagram for known ultracool dwarfs, using the same
    colors and symbols as Figure~\ref{fig.iy.iz}; in addition, early-T dwarfs
    are indicated by dark blue triangles, and mid- and late-T dwarfs by light
    blue triangles.  The vertical dotted line indicates our \zy\ cut, which we
    applied only to objects with $\sigma_z<0.2$~mag and at least two \zps\
    detections.  The horizontal dashed line repesents our \wbwc\ cut, which we
    applied to all objects in our search.  We selected objects below and to the
    right of these lines.}
\label{fig.w2w3.zy}
\end{center}
\end{figure}

\begin{figure}
\begin{center}
  \includegraphics[width=1.00\columnwidth, trim = 20mm 0 9mm 0]{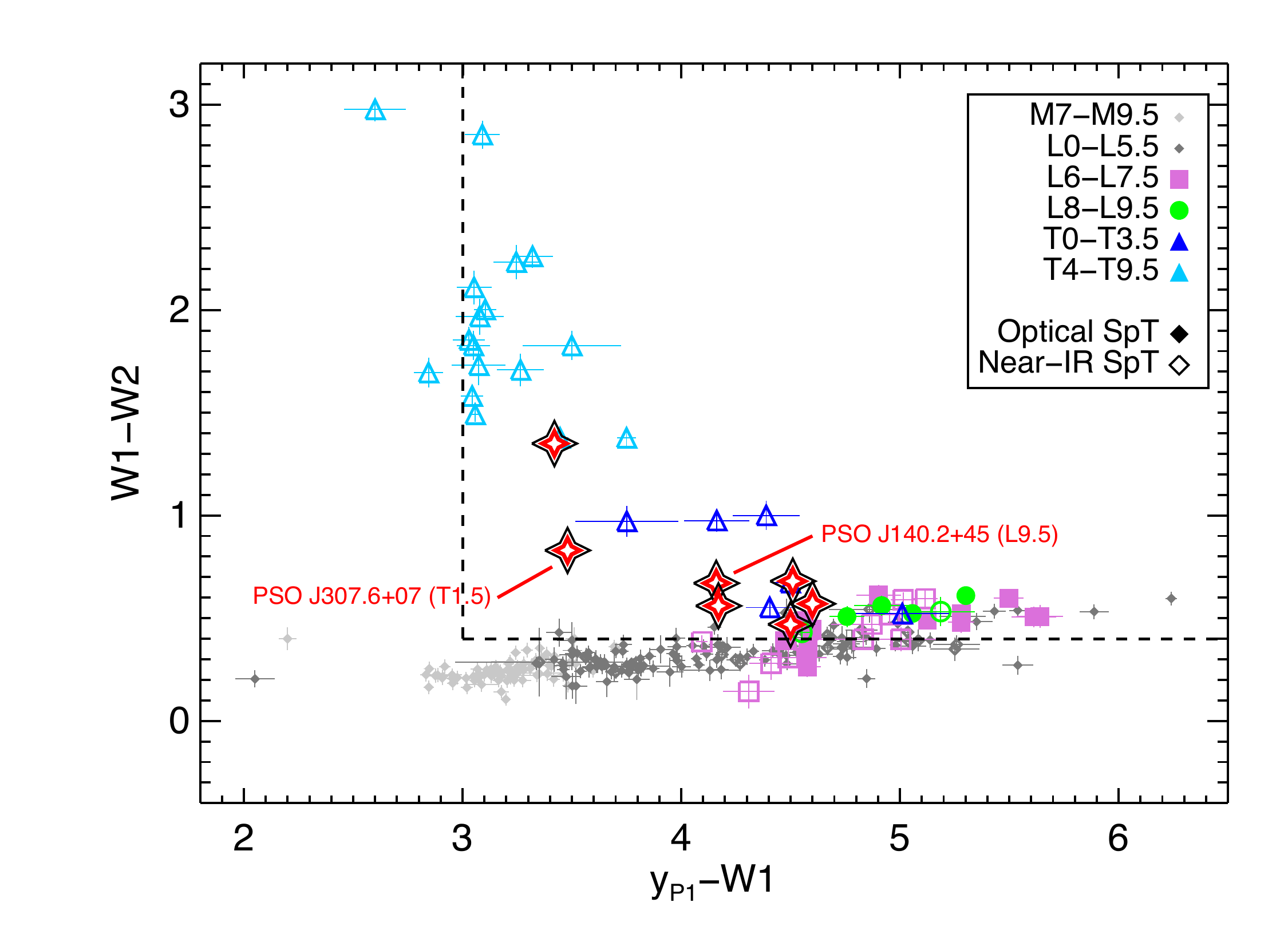}
  \caption{$W1-W2$ vs. $y_{\rm P1}-W1$ diagram for known ultracool dwarfs, using
    the same colors, lines and symbols as Figures~\ref{fig.iy.iz} and
    \ref{fig.w2w3.zy}.  We selected objects with colors above and to the right
    of the dashed lines.  All of our discoveries lie in the typical region of
    this color space for late-L and early-T dwarfs, except that PSO~J307.6+07 is
    fairly blue in $y_{\rm P1}-W1$ for an early-T dwarf.}
\label{fig.w1w2.yw1}
\end{center}
\end{figure}

To extract candidate L/T transition dwarfs from our PS1+\WISE\ database, we
applied the following quality-of-detection criteria (items shown in parentheses
refer to flags within the database):
\begin{enumerate}
\item Detected in at least two separate \yps\ frames ({$\tt y\!:\!nphot\ge2$}), to
  reject transients.
\item $\sigma_y<0.2$~mag ({$\tt y\!:\!err<0.2$}), establishing S/N~$>5$ as the
  detection threshold for \yps.
\item ``Good'' or ``poor'' quality \yps\ photometry, but not ``bad'', as defined
  in the PS1 DVO database ({$\tt y\!:\!flags=256$} or {$\tt y\!:\!flags=512$}),
  i.e., clear PSF identification, no saturated objects or cosmic rays.
\item $W1$ and $W2$ detections have $\mathrm{S/N}>2$ ({\tt ph\_qual = A, B,} or
  {\tt C}).
\item $W1$ and $W2$ detections are not saturated ({$\tt w1sat=0$} and {$\tt
    w2sat=0$}).
\item $W1$ and $W2$ detections are point sources ({$\tt ext\_flg=0$}).
\item $W1$ and $W2$ detections are unlikely to be variable ({$\tt
    var\_flg\le4$}).
\end{enumerate}
We thus required that candidates have good detections in \yps, $W1$, and $W2$,
but not necessarily in any other bands.  We then applied the following color
criteria, which are illustrated by Figures~\ref{fig.iy.iz}--\ref{fig.w1w2.yw1}:
\begin{enumerate}
\item No more than one total detection in either \gps\ or \rps\ ({$\tt
    g\!:\!nphot+r\!:\!nphot<2$}), unless each band had only a single detection
  with $\sigma>0.2$~mag.  Among known L/T transition dwarfs, none are detected
  in \gps\ and only one in \rps\ --- the nearby, unusually blue L6 dwarf
  SDSS~J1416+1348 \citep{Bowler:2010gd}.  We therefore expect previously
  undiscovered L/T transition dwarfs to be too red to show up in \gps\ and
  \rps\, and this criterion rejects objects that are clearly detected in these
  bluer bands.
\item \iz\ $\ge1.8$~mag (Figure~\ref{fig.iy.iz}), to exclude objects of spectral
  type mid-L and earlier.  We applied this criterion only when the \ips\ and
  \zps\ photometry for an object met the same quality standards required for
  \yps\ detections, i.e., $\sigma_i<0.2$~mag and $\sigma_z<0.2$~mag ({$\tt
    i\!:\!err<0.2$} and {$\tt z\!:\!err<0.2$}) with at least two detections in
  both \ips\ and \zps\ ({$\tt i\!:\!nphot\ge2$} and {$\tt z\!:\!nphot\ge2$}, to
  reject transient detections).  Of the L/T transition dwarfs too faint to have
  been previously discovered in 2MASS and SDSS, even the brightest are likely to
  be near the detection limits of \ips\ and \zps.  (In fact, we do not detect
  any known T dwarfs in \ips.)  So we only applied this cut when we had good
  quality photometry in both bands, and accepted objects with marginal or null
  detections in \ips\ and \zps\ as long as they passed the other cuts.  We note
  that this criterion is somewhat more relaxed than the $i-z>2.2$~mag and
  $\sigma_z<0.12$~mag cuts used by \citet{Chiu:2006jd} in their SDSS brown dwarf
  search.
\item \iy\ $\ge2.8$~mag (Figure~\ref{fig.iy.iz}), also to exclude objects of
  spectral type mid-L and earlier.  Again, we only applied this criterion if the
  \ips\ photometry for an object met the same quality standards required for
  \yps\ detections, i.e., $\sigma_i<0.2$~mag ({$\tt i\!:\!err<0.2$}) and
  detection in \ips\ in at least two separate frames ({$\tt i\!:\!nphot\ge2$}).
\item \zy\ $\ge0.6$~mag (Figure~\ref{fig.w2w3.zy}), to screen out early-M
  dwarfs.  Similarly to the \iz\ and \iy\ cuts, we only applied this cut if the
  \zps\ photometry for an object met the same quality standards required for
  \yps\ detections, i.e., $\sigma_z<0.2$~mag ({$\tt z\!:\!err<0.2$}) and
  detection in \zps\ in at least two separate frames ({$\tt z\!:\!nphot\ge2$}).
  This cut was also used by D11 in their PS1+2MASS T~dwarf search.
\item \ywa\ $\ge3.0$~mag (Figure~\ref{fig.w1w2.yw1}), to screen
  out early- and mid-M dwarfs.  We expect this cut will have also rejected some
  late T~dwarfs (SpT $\gtrsim$ T6), which have bluer \ywa\ colors than L/T
  transition dwarfs.
\item \wawb\ $\ge0.4$~mag (Figure~\ref{fig.w1w2.yw1}), to remove
  as many M and early-L dwarfs as possible.  \citet{Kirkpatrick:2011ey} and M13
  used this same cut in their \WISE\ search for nearby L and T dwarfs, but also
  restricted their search to bright objects with no 2MASS counterpart (i.e.,
  objects which had moved more than 3'' between 2MASS and \WISE), criteria that
  we do not use here.  \citet{Liu:2011hc} used a redder $W1-W2>0.7$~mag cut to
  search only for T dwarfs.
\item \wbwc\ $\le2.5$~mag (Figure~\ref{fig.w2w3.zy}), to screen
  out extragalactic sources, as suggested by Figure~12 of \citet{Wright:2010in}.
  Many of our candidates were not actually detected in $W3$, but \WISE\ reports
  a lower magnitude limit for all non-detections, so we used the \wbwc\ value as
  an upper limit in those cases.  \citet{Kirkpatrick:2011ey} and M13 used
  $W1-W2>0.96(W2-W3)-0.96$~mag for this purpose.
\end{enumerate}
We then applied cuts based on spatial position:
\begin{enumerate}
\item We rejected all candidates within $3^\circ$ of the Galactic plane.
\item To avoid objects in reddened star forming regions, we inspected the (RA,
  Dec) positions of candidates within $20^\circ$ of the Galactic plane and
  rejected those that were clumped in tight groups.
\item We further rejected all remaining candidates in the highly reddened
  regions identified by \citet{Cruz:2003fi}, unless a candidate had a proper
  motion measured with ${\rm S/N}>3$ (based on 2MASS and PS1 astrometry).
\end{enumerate}

We then reviewed the available images for each remaining candidate in all five PS1
filters ({\it g, r, i, z, y}), all three 2MASS filters ({\it J, H, K}), and
\WISE\ $W1$, $W2$, and $W3$.  We rejected objects that were clearly artifacts
(e.g., diffraction spikes or halos) in one or more bands or non-point sources
such as galaxies.

\begin{figure}
\begin{center}
  \includegraphics[width=1.00\columnwidth, trim = 10mm 0 5mm 0]{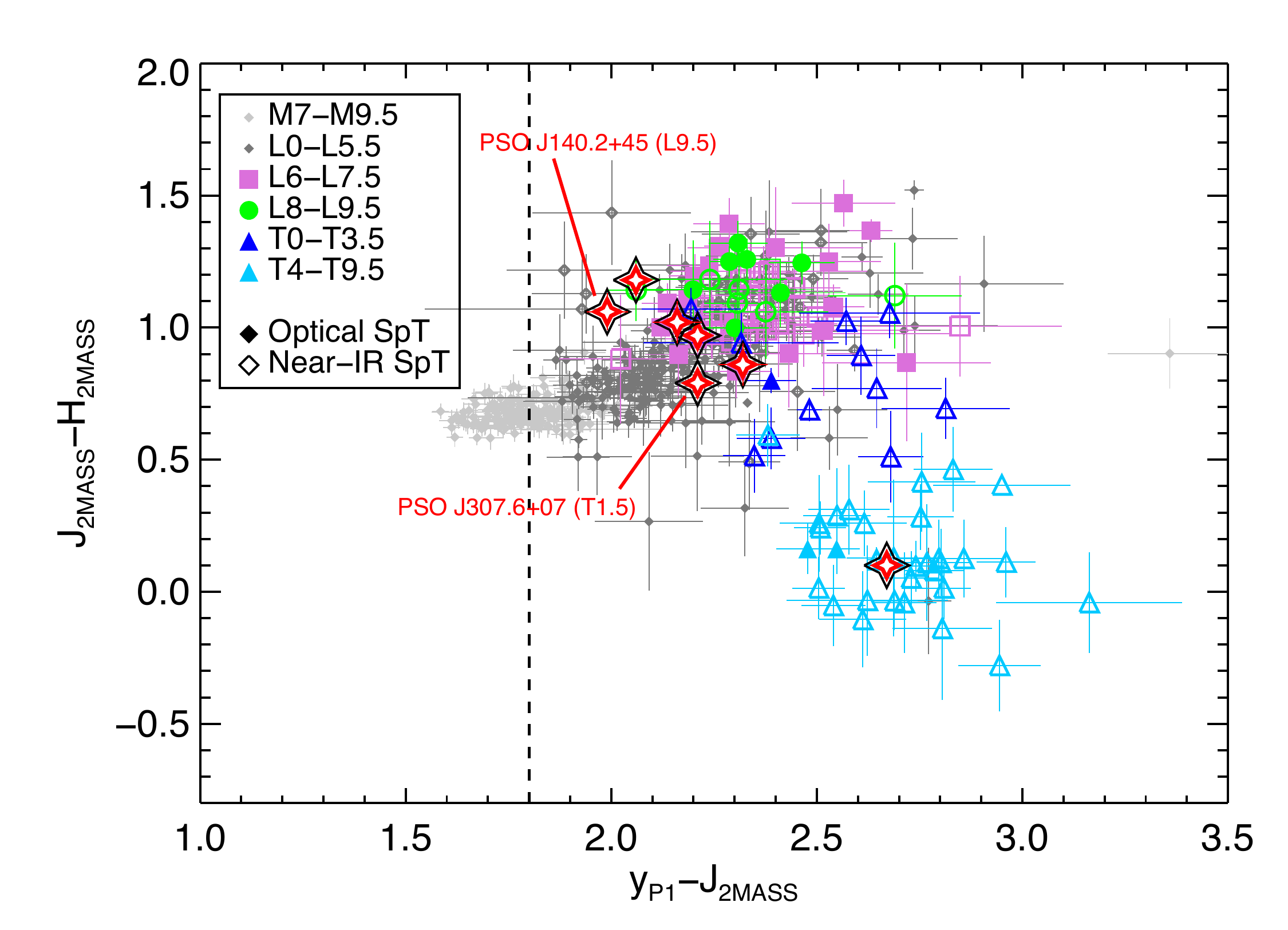}
  \caption{$J_{\rm 2MASS}-H_{\rm 2MASS}$ vs. $y_{\rm P1}-J_{\rm 2MASS}$ diagram
    for known ultracool dwarfs, using the same colors, lines and symbols as
    Figures~\ref{fig.iy.iz} and \ref{fig.w2w3.zy}.  We selected objects to the
    right of the dashed line.  The red object at lower right is the T5
    PSO~J339.0+51; the other red objects have spectral types L9$-$T1.
    PSO~J339.0+51 has typical $J_{\rm 2MASS}-H_{\rm 2MASS}$ and $y_{\rm
      P1}-J_{\rm 2MASS}$ values for its spectral type, while all of our other
    new discoveries are blue in $y_{\rm P1}-J_{\rm 2MASS}$ compared to known
    late-L and early-T dwarfs, possibly because previous searches have tended to
    find redder objects.}
\label{fig.JH.yJ}
\end{center}
\end{figure}

As a final filter, we required objects to have sufficiently red $y-J$ colors to
be in the L/T transition.  We cross-matched our candidate list with
  the 2MASS, UKIDSS DR8 \citep{Hambly:2008kr,Lawrence:2012wh}, and VISTA DR1
  \citep{Cross:2012jz} catalogs, and used $J$ magnitudes from those catalogs
  when available with $\sigma_J<0.1$~mag.  The UKIDSS photometric system uses
  MKO standards \citep{Tokunaga:2002ex} and is described in
  \citet{Hewett:2006hy}.  The VISTA filters are similar to those of UKIDSS,
  except that VISTA uses a $K_s$ filter similar to that of 2MASS
  \citep{Lodieu:2012go}\footnote{For VISTA bandpass information, see
    http://casu.ast.cam.ac.uk/surveys-projects/vista/technical/filter-set.}.  (A
  precise calibration of the VISTA photometric system has not yet been
  published.)  For objects not found in any of these catalogs, we obtained our
own photometry with UKIRT/WFCAM on the MKO system (as described in
Section~\ref{obser}).  We selected objects with $y_{\rm P1}-J_{2MASS}\ge1.8$~mag
(Figure~\ref{fig.JH.yJ}) or $y_{\rm P1}-J_{MKO}\ge1.9$~mag,
removing bluer objects that were likely to be late-M and early-L dwarfs from our
candidate list.  This cut does permit objects bluer than the $2.2<y_{\rm
  P1}-J_{2MASS}<5.0$~mag cut used by D11, who searched exclusively for T dwarfs.
Unlike D11, we did not apply a $J-H$ cut to our sample.

Because of the spectral changes endemic to L/T objects, \yps\ absolute
magnitudes are almost flat across the transition, from about L8 to T3 spectral
types.  Therefore, choosing a \yps-magnitude limit approximately creates a
volume-limited sample, with only modest contamination by intrinsically brighter
objects that lie beyond the search radius.  For follow-up observations, we chose
candidates with $y_{\rm P1}<19.3$~mag to search the solar neighborhood
for single L/T objects out to 25~pc, acknowledging that our search
  will also detect unresolved binaries out to larger distances.

\section{Observations}
\label{obser}

We obtained follow-up near-IR imaging of our candidates using WFCAM
\citep{Casali:2007ep} on the United Kingdom InfraRed Telescope (UKIRT) in queue
mode.  Over five nights spanning 2012 September 19--21 and 2012 December 13--14
UT, we observed a total of 308 candidates in different combinations of the $Y$,
$J$, $H$ and $K$ bands (MKO system).  Conditions were generally clear with
seeing $\approx0.6''$ on the September nights, while the December nights had
many high clouds with seeing $\approx0.8''$.  Data were reduced and calibrated
at the Cambridge Astronomical Survey Unit using the WFCAM survey pipeline
\citep{Irwin:2004ej,Hodgkin:2009jr}.

We obtained low resolution near-IR spectra for our seven discoveries over eight
nights in 2012 September, October and November and 2013 January with the NASA
Infrared Telescope Facility (IRTF).  We obtained additional spectra
  for PSO~J307.6784+07.8263 (hereinafter PSO~J307.6+07) on 2013 April 3--5,
  contemporaneously with the nearby M1V star 2MASS~J20410101+0014278
  \citep[][hereinafter 2MASS~J2041+0014]{West:2011dx} for comparison.  We used
the facility spectrograph SpeX \citep{Rayner:2003hf} in prism ($R\sim100$) and
SXD ($R\sim750$) modes with the $0.5''$ and $0.8''$ slits.  Details of our
observations are listed in Table~\ref{tbl1}.  We observed a nearby A0V star
contemporaneously with each science target for telluric calibration.  All
spectra were reduced using version 3.4 of the Spextool software package
\citep{Vacca:2003fw,Cushing:2004bq}.

Initial spectral typing of candidates was accomplished using spectral indices
from \citet{Burgasser:2006cf}, with spectral types assigned using the polynomial
fits from \citet{Burgasser:2007fl}.  The spectra were then visually compared to
near-infrared spectra for L dwarf \citep{Kirkpatrick:2010dc} and T dwarf
spectral standards \citep{Burgasser:2006cf} for final assignment of spectral
types.  Following the procedure of \citet{Kirkpatrick:2010dc}, when comparing
our candidate spectra with L dwarf standards, we first compared only the
$0.9-1.4\ \mu$m portions of the spectra to evaluate the goodness of fit.  If the
best match for a candidate was determined to be an L dwarf standard, then we
compared the candidate spectrum to that of the standard over $1.4-2.4\ \mu$m to
determine if the candidate was unusually red or blue for its spectral type.
When comparing candidates to T dwarf standards, we judged the goodness of fit
for the entire $0.9-2.4\ \mu$m window simultaneously.

\section{Results}
\label{results}
\begin{figure}
\begin{center}
  \includegraphics[width=0.83\columnwidth]{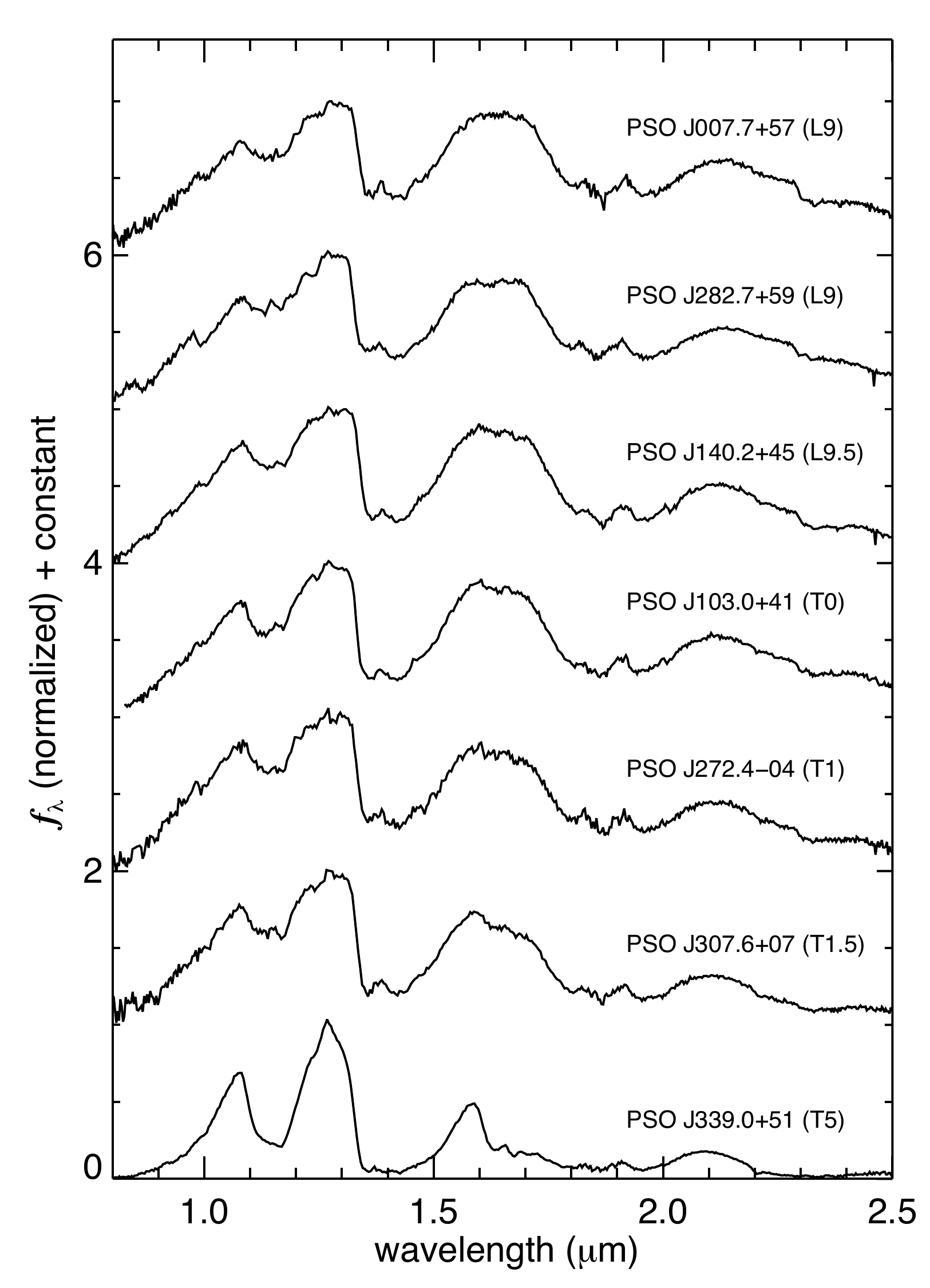}
  \caption{SpeX prism spectra for our seven objects, normalized at the $J$-band
    peak ($1.27\,\mu$m), arranged from earliest to latest spectral type and
    offset by a constant.  Spectral typing was done by visual comparison with
    the near-infrared standards defined by \citet{Burgasser:2006cf} and
    \citet{Kirkpatrick:2010dc}.}
\label{fig.7stack}
\end{center}
\end{figure}

We present here seven bright initial L/T transition discoveries from our ongoing
search.  Our photometric distances place all seven within $9-15$~pc of the Sun,
assuming they are single objects and not unresolved binaries.  The
astrometric, photometric, and spectral properties of these objects are listed in
Tables~\ref{tbl2} and \ref{tbl3}.  Near-IR SpeX prism spectra for these objects
are shown in Figure~\ref{fig.7stack}.

Four of our seven new objects have been announced by M13 in their search for
late-T and Y dwarfs in \WISE; our discovery of all four was independent (our
spectroscopic confirmation preceded publication of their paper).  For the most
part, we agree with the spectral indices and types published by M13, and we
present additional photometry, photometric distances, astrometry, and spectral
indices for the four objects, as well as higher-resolution SXD spectra for two
of them.  Specific objects are discussed in Section~\ref{discuss}.

We used the $W2$ absolute magnitude vs. spectral type polynomial from
\citet{Dupuy:2012bp}\footnote{Updated polynomials can be found in the Database
  of Ultracool Parallaxes maintained by Trent Dupuy at
  https://www.cfa.harvard.edu/\mytilde
  tdupuy/plx/Database\_of\_Ultracool\_Parallaxes.html.  Here we use the version
  posted on 2012 June 09.} to calculate photometric distances.  The $W2$
polynomial was determined using objects of spectral types M5.5 to T9 and has an
rms scatter about the fit of 0.35~mag (slightly lower than the rms scatter of
0.39~mag for the $W1$ polynomial).  However, the scatter about the fit is
smaller for later spectral types than for M dwarfs.  \citet{Liu:2011hc}
calculated an rms scatter of 0.18~mag for a $W2$ polynomial relationship using
only objects with spectral types L5 and later.  We adopted this rms to determine
the uncertainty in the distance.

We combined 2MASS and PS1 astrometry to calculate proper motions for our
discoveries and used our photometric distances to estimate tangential
velocities.  A comparison of these $v_{tan}$ estimates to the $\sigma_{v_{tan}}$
vs. $v_{tan}$ relations in \citet[][Figure 31]{Dupuy:2012bp} indicates that all seven
discoveries are very likely to be members of the thin disk.

Figures~\ref{fig.iy.iz}--\ref{fig.JH.yJ} display \ips\ through $W3$ colors
of known ultracool dwarfs\footnote{Compiled from DwarfArchives.org as of April
  2011, \citet{Leggett:2010cl}, \citet{Burgasser:2011dr}, and
  \citet{Gelino:2011cw}.}, overlaid with our seven discoveries.  Compared with
other early-T dwarfs, PSO~J307.6+07 (T1.5) is unusually blue in $y_{\rm P1}-W1$
(by $\sim0.7$~mag) and \yps$-J_{\rm 2MASS}$ (by $\sim0.3$~mag), suggesting that
its \yps\ flux is unusually bright.  In addition, the six objects that have
spectral types L9$-$T1.5 are all bluer than typical late-L and early-T dwarfs by
about half a magnitude in \yps$-J_{\rm 2MASS}$.  We find no systematic reason
why our search method would preferentially discover objects that are bluer in
\yps$-J_{\rm 2MASS}$.  It is possible that the previous searches that identified
the known objects were biased toward finding objects with redder \yps$-J_{\rm
  2MASS}$ colors (e.g., D11).

\section{Discussion}
\label{discuss}

\subsection{Candidate Spectroscopic Variables}
\label{specvar}
\begin{figure}
\begin{center}
  \includegraphics[width=0.99\columnwidth]{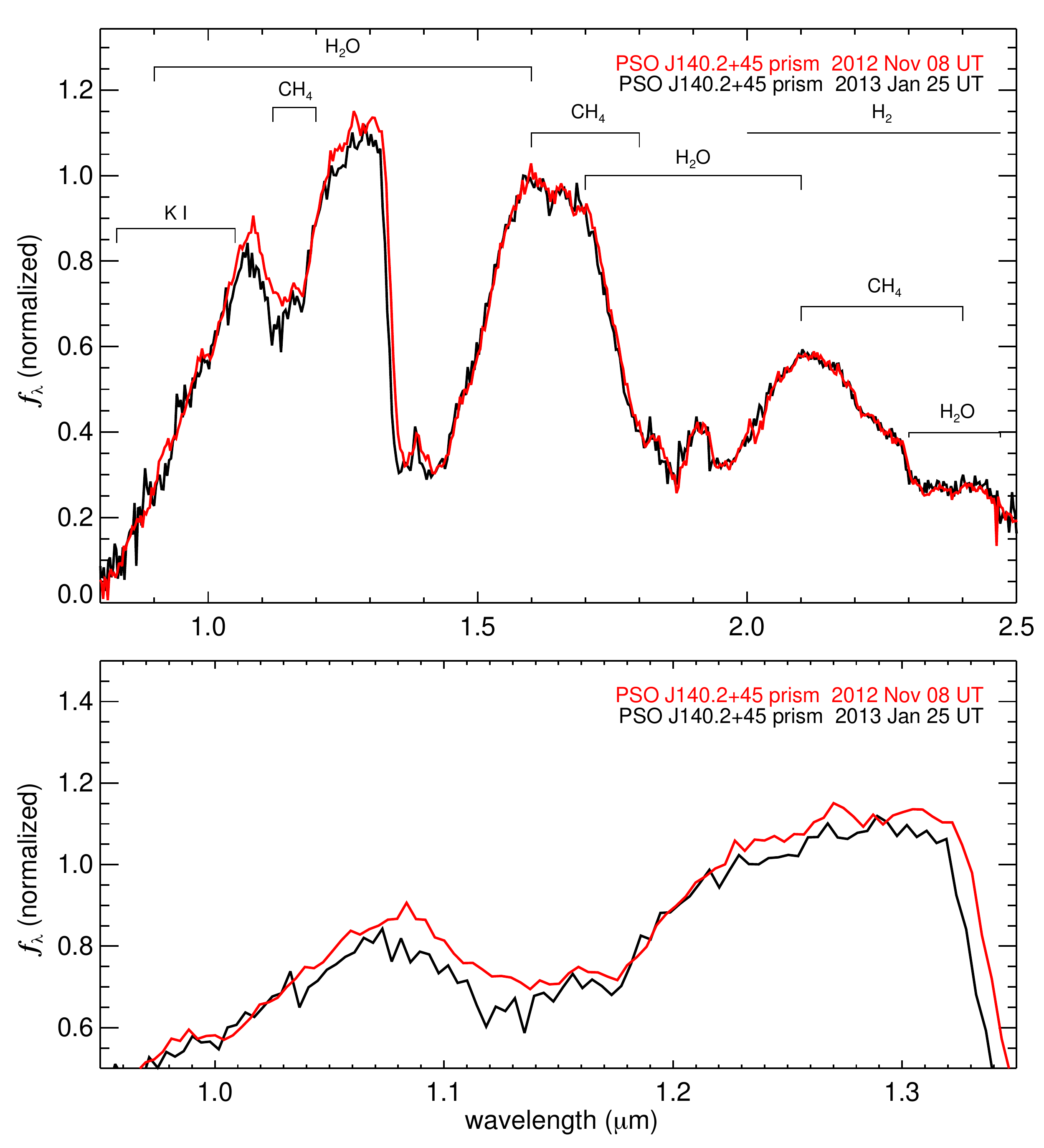}
  \caption{{\it Top:} Prism spectra ($R\sim100$) for PSO~J140.2+45 taken in
    November 2012 (black) and January 2013 (red), both with SpeX on IRTF.  {\it
      Bottom:} Same as top window, but showing only the $Y$- and $J$-bands.  The
    changes in flux in the $Y$-band peak ($1.06-1.10\ \mu$m) and marginally in
    the $J$-band ($1.16-1.33\ \mu$m) suggest variability similar to that
    previously detected in two other early-T dwarfs
    \citep{Artigau:2009bk,Radigan:2012ki,Apai:2013fn}.  The spectra have been
    normalized to the $H$-band peak ($1.58\,\mu$m) to highlight the $Y$- and
    $J$-band variations.}
\label{fig.5-011389.zoom}
\end{center}
\end{figure}

PSO~J140.2308+45.6487 (hereinafter PSO~J140.2+45), also known as WISE~0920+4538,
and PSO~J307.6+07, also known as WISE~J2030+0749, are the second
  and third L/T transition dwarfs to be identified as candidate
  near-IR variables via spectroscopy, following SDSS
  J125453.90$-$012247.4 \citep{Goldman:2008es}.  Spectra of PSO~J140.2+45
taken on two different nights (Figure~\ref{fig.5-011389.zoom}) show small but
clear differences in the $Y$- and $J$-bands, while for PSO~J307.6+07
(Figure~\ref{fig.11-022304.zoom}) the difference in $J$-band is readily
apparent.  For comparison, Figure~\ref{fig.12-014644} shows SXD and prism
spectra for PSO~J339.0734+51.0978 (hereinafter PSO~J339.0+51) taken ten nights
apart with no clear variability.

For all three of these objects, we have normalized the pairs of
  spectra to the $H$-band peak ($1.58\,\mu$m).  These spectra are not absolutely
  calibrated, so we cannot determine with certainty the absolute changes in flux
  in each band; however, we note that each object's $H$- and $K$-bands show no
  significant change in spectral shape or in flux relative to the adjacent
  absorption bands.  We calculated the differences in flux in the $Y$- and
  $J$-bands for each pair of spectra by first convolving our spectra in the
  wavelength intervals $0.95-1.13\ \mu$m and $1.16-1.33\ \mu$m with the UKIDSS
  $Y$ \citep[$0.97-1.07\ \mu$m,][]{Hewett:2006hy} and $J$ \citep[$1.17-1.33\
  \mu$m,][]{Tokunaga:2002ex} filter profiles, respectively.  We then integrated
  the convolved spectra over each interval, subtracted to get the differences in
  flux, and converted these flux differences and their uncertainties to
  magnitudes.  In addition, we calculated the differences in magnitude in the
  narrow peaks of the $Y$-band ($1.06-1.10\ \mu$m) and $J$-band ($1.27-1.30\
  \mu$m) using the same method.

Details of each candidate variable are described below; here we
discuss briefly the possible causes of variability.  \citet{Artigau:2009bk} and
\citet{Radigan:2012ki} explain the $J$-band photometric variations in the
early-T dwarfs SIMP~0136+0933 and 2MASS~2139+0220 as consequences of the
clearing of atmospheric condensate clouds, the process thought to cause the
observed $J$-band brightening in the L/T transition.  The variations are seen
only in the $J$-band because the condensates in question are primarily silicates
that absorb most strongly at $J$-band wavelengths compared to other (gaseous)
opacity sources \citep{Ackerman:2001gk,Burrows:2006ia}.  The clearing process is
not well understood.  Several scenarios have been proposed wherein condensate
clouds thin gradually (reducing their opacities), rain out suddenly, or break up
\citep[e.g.,][]{Ackerman:2001gk,Knapp:2004ji,Tsuji:2005cd,Burrows:2006ia,Marley:2010kx}.
Models using a mixture of thin and thick clouds may generate a better match to
observed early-T dwarf spectra \citep{Buenzli:2012gd}.  \citet{Radigan:2012ki}
attempt a variety of model atmosphere fits to their observations of 2MASS
2139+0220 and find success combining models with different condensate
sedimentation efficiences, interpreted as heterogenous cloud formations.
$J$-band variation over a period of several days could therefore be a sign of
thinner and thicker clouds or even breaks in clouds, changing due to rotation
and/or convective processes.

\subsubsection{PSO~J140.2308+45.6487 (WISE~092055.40+453856.3)}
\label{140discuss}
PSO~J140.2+45 was first identified photometrically as a candidate L4--L5 brown
dwarf (WISE~0920+4538) by \citet{Aberasturi:2011bk} based on \WISE/2MASS/SDSS
photometry and measurable proper motion.  Our spectra of PSO~J140.2+45 taken two
and a half months apart (Figure~\ref{fig.5-011389.zoom}) indicate a
  marginal decrease of $0.05\pm0.04$~mag in the full $Y$-band, but a signficant
  decrease of $0.14\pm0.03$~mag in the $Y$-band peak.  The decrease in $J$-band
  is also marginal: $0.05\pm0.02$~mag in the full band, and $0.03\pm0.02$~mag in
  the $J$-band peak.  Visual classification of our SpeX spectrum for PSO
J140.2+45 gives a spectral type L9.5, consistent with the L9.6 spectral type
derived from the spectral indices.  M13 classify this object as L9 and a weak
binary candidate.  Further observations are needed to confirm the
  near-IR variability of this object.

\subsubsection{PSO~J307.6784+07.8263 (WISE~203042.79+074934.7)}
\label{307discuss}
Figure~\ref{fig.11-022304.sxd} compares our SXD spectrum for PSO~J307.6+07 with
that of the T2 standard SDSS~1254$-$0122
\citep{Leggett:2000ja,Cushing:2005ed,Burgasser:2006cf}.  No unusual spectral
features are apparent in PSO~J307.6+07 at this resolution ($R\sim750$).  The
spectrum of PSO~J307.6+07 is reasonably close to that of SDSS~1254$-$0122 but
has slightly weaker methane absorption at $\sim$1.15 $\mu$m.  The spectral
indices have a mean classification of T$0.9\pm1.3$.  Given these estimates, we
classify PSO~J307.6+07 as a T1.5 dwarf.  M13 identified this object as
WISE~2030+0749 and assigned it the same spectral type.

\begin{figure}
\begin{center}
  \includegraphics[width=0.99\columnwidth]{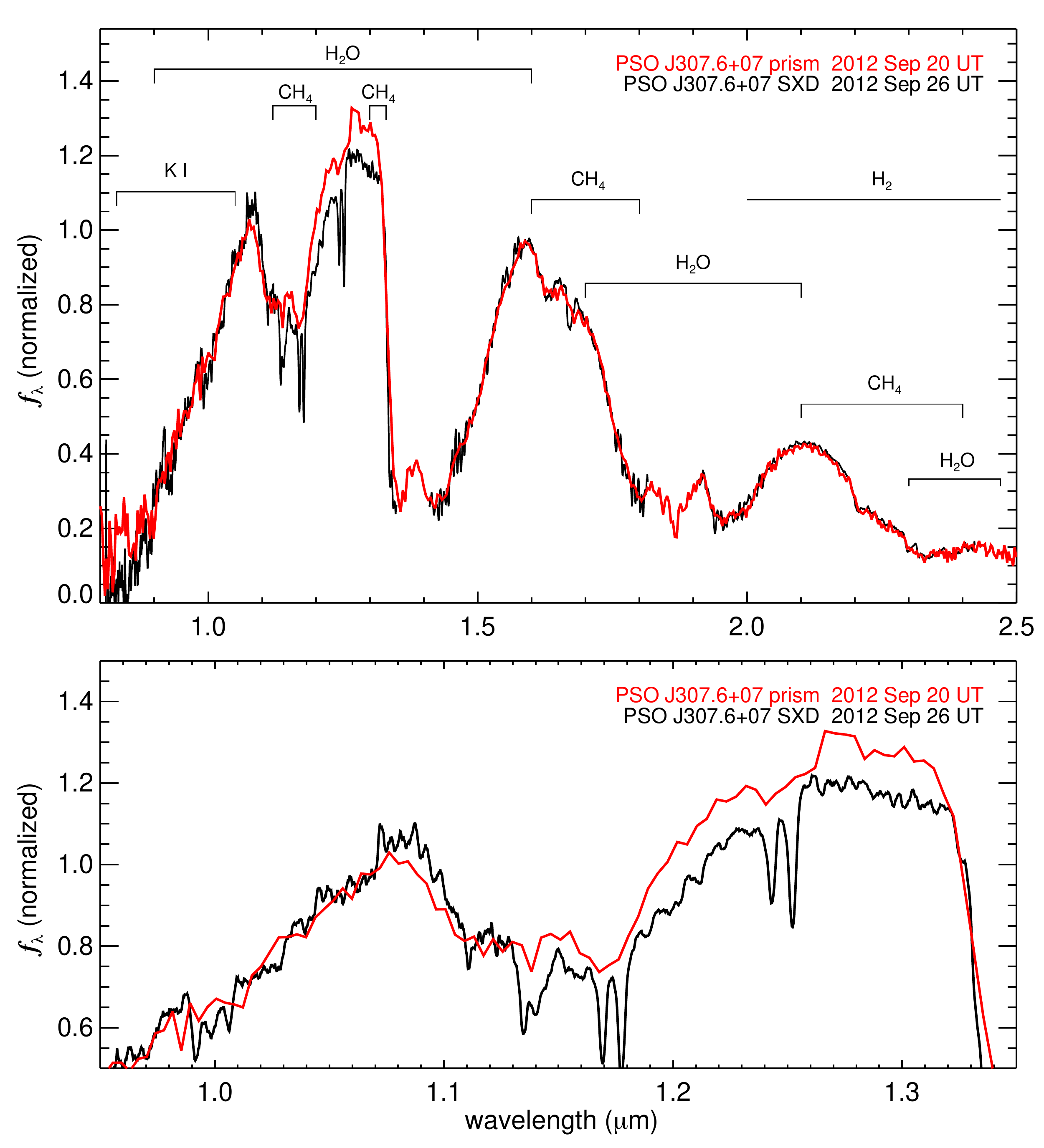}
  \caption{{\it Top:} Spectra for PSO~J307.6+07 taken six nights apart, both
    with SpeX on IRTF.  The prism spectrum (red) has $R\sim100$ while the SXD
    spectrum (black) has $R\sim750$ and is smoothed by an 8-pixel box.  {\it
      Bottom:} Same as top window, but showing only the $Y$- and $J$-bands.  The
    change in flux in the $J$-band ($1.16-1.33\ \mu$m) clearly suggests
    variability.  The slight difference visible in the $Y$-band ($1.00-1.13\
    \mu$m) flux is not statistically significant.  The spectra have been
    normalized to the $H$-band peak ($1.58\,\mu$m) to highlight the $J$-band
    variation.}
\label{fig.11-022304.zoom}
\end{center}
\end{figure}

\begin{figure}
\begin{center}
  \includegraphics[width=0.99\columnwidth]{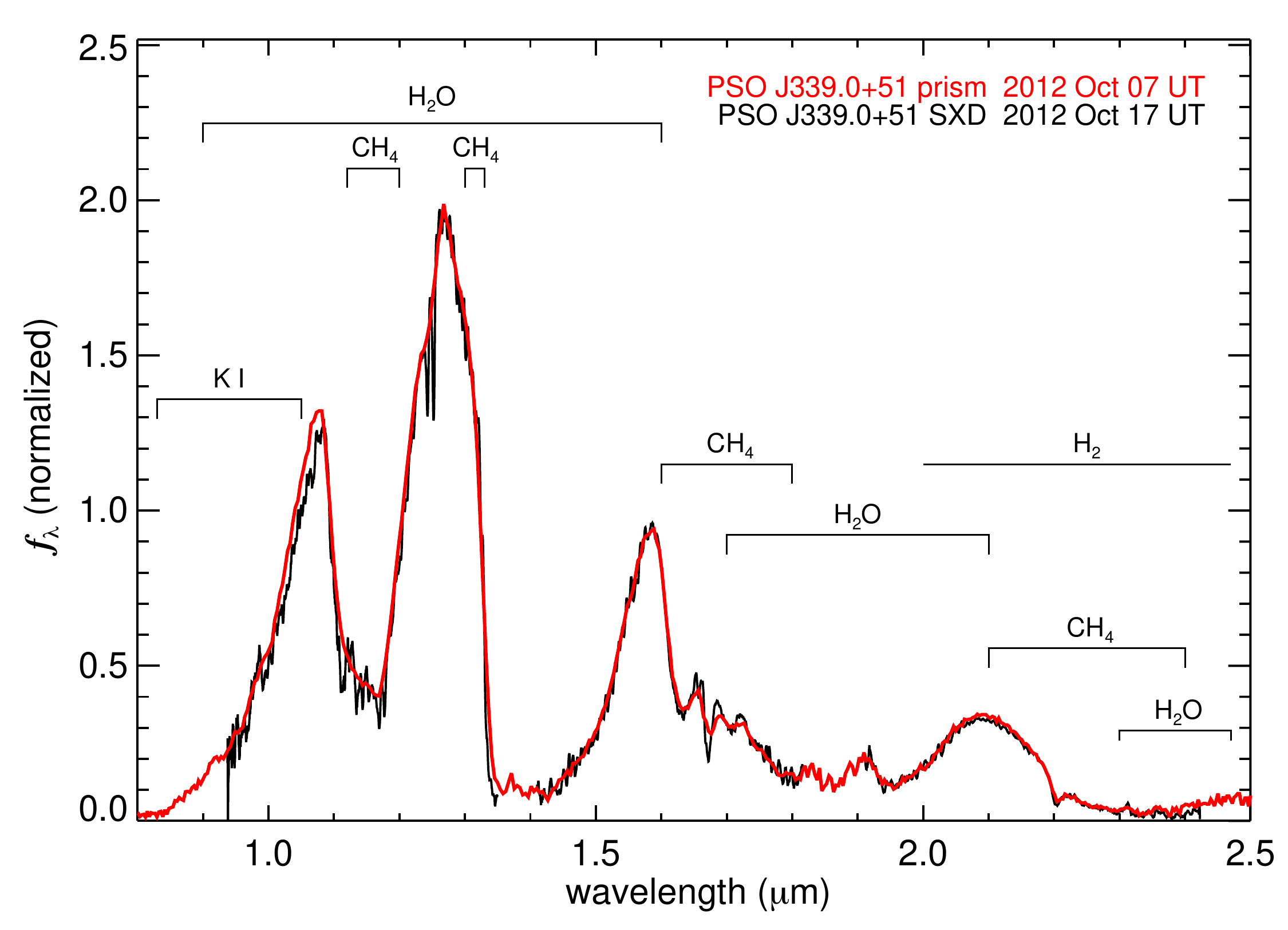}
  \caption{Spectra for PSO~J339.0+51 taken ten nights apart, both with SpeX on
    IRTF.  The prism spectrum (red) has $R\sim100$ while the SXD spectrum
    (black) has $R\sim750$ and is smoothed by an 8-pixel box.  No clear sign of
    variability is seen between these two spectra; the slight difference visible
    in the $Y$-band ($1.00-1.13\ \mu$m) flux is not statistically significant.
    As in Figures~\ref{fig.5-011389.zoom} and \ref{fig.11-022304.zoom}, these
    spectra have been normalized to the $H$-band peak.}
\label{fig.12-014644}
\end{center}
\end{figure}

\begin{figure}
\begin{center}
  \includegraphics[width=0.85\columnwidth]{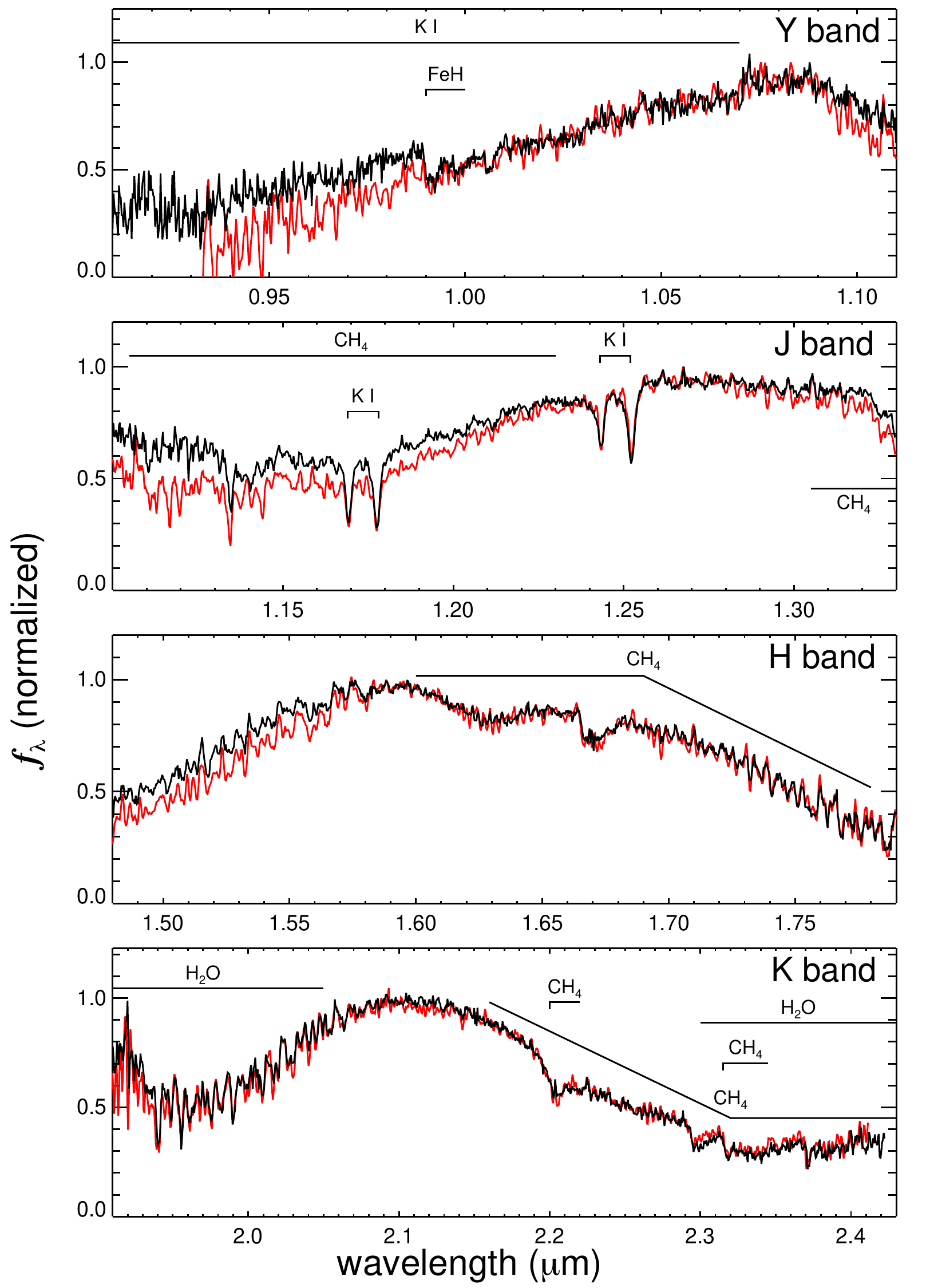}
  \caption{SpeX SXD spectra for PSO~J307.6+07 (T1.5, black) and the T2 standard
    SDSS~1254$-$0122 \citep[red,][]{Cushing:2005ed}.  The spectra are normalized
    to the peak value in each band and show the same prominent absorption
    features.  The relative faintness of SDSS~1254$-$0122 in parts of the $Y$-,
    $J$- and $H$-bands is expected because PSO~J307.6+07 is half a spectral type
    earlier.}
\label{fig.11-022304.sxd}
\end{center}
\end{figure}

Our spectra of PSO~J307.6+07 taken six nights apart in 2012 September
(Figure~\ref{fig.11-022304.zoom}) show a decrease of $0.11\pm0.02$~mag in the
full $J$-band, and $0.10\pm0.02$~mag in the $J$-band peak.  The overlaid spectra
also suggest an increase in the $Y$-band flux over the six-night span, but we
calculate this difference as only $0.02\pm0.06$~mag in the full $Y$-band, and
$0.02\pm0.05$~mag in the $Y$-band peak.

\begin{figure}
\begin{center}
  \includegraphics[width=0.99\columnwidth]{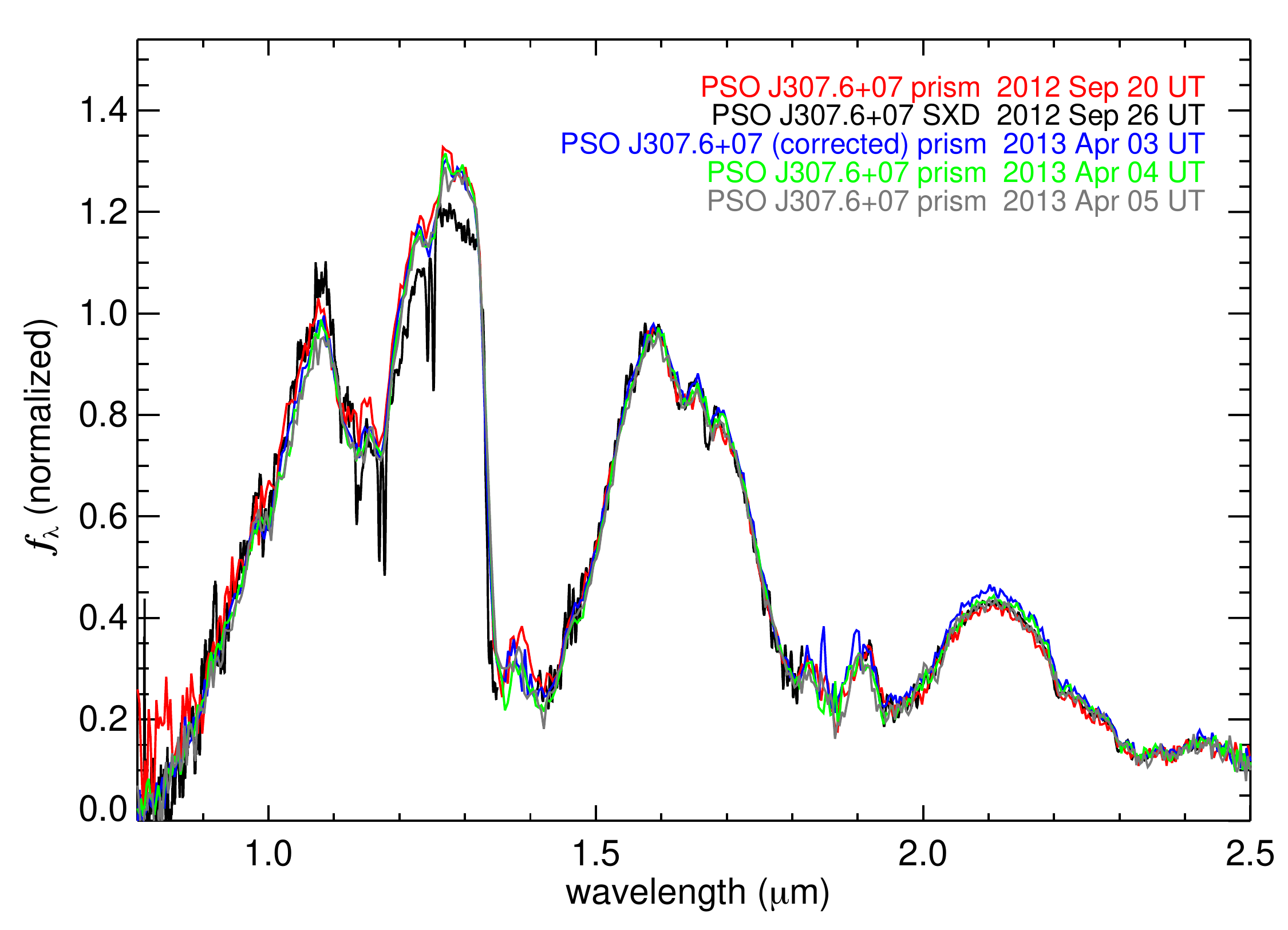}
  \caption{Spectra of PSO~J307.6+07 taken in 2012 September (red and black, from
    Figure~\ref{fig.11-022304.zoom}) and 2013 April (blue, green, and grey).
    The September 26 SXD spectrum (black) is smoothed by an 8-pixel box.  The
    April 3 spectrum (blue) has been corrected for dimming seen in the
    comparison M1V star 2MASS~J2041+0014 on that night.  As in previous figures,
    these spectra have been normalized to the $H$-band peak.  Unlike the
    September spectra, the April spectra show no variation.  In the $Y$-band,
    both of the September spectra differ from the April spectra; in the
    $J$-band, only September 26 is different.}
\label{fig.5epoch}
\end{center}
\end{figure}

Figure~\ref{fig.5epoch} shows spectra of PSO~J307.6+07 obtained on three
consecutive nights in 2013 April, along with the two 2012 September spectra from
Figure~\ref{fig.11-022304.zoom}.  When examining the spectra of the comparison
M1V star 2MASS~J2041+0014 (presumably a non-variable source) from the same three
April nights, we noticed a dimming at wavelengths $\lesssim1.3\mu$m on April 3
compared to the two subsequent nights.  This dimming may be due to the presence
of a crescent Moon about 6$^\circ$ from the standard star during this night's
observations.  To correct for the dimming, we divided the April 4 spectrum of
2MASS~J2041+0014 by its April 3 spectrum, and then multiplied this quotient into
the April 3 spectrum of PSO~J307.6+07.  With this correction applied, the April
data for PSO~J307.6+07 show essentially no variation over the three nights.  In
the $J$-band, the April spectra also match well the 2012 September 20 spectrum,
with only the September 26th SXD spectrum showing a clear difference in flux.
However, in the $Y$-band, the April spectra all have noticeably lower flux than
both of the September spectra.

Collectively, our data show significant spectrospcopic variation for
PSO~J307.6+07 in 2012 September, and no variation in 2013 April.  The reason for
this apparent contradiction is unclear.  The long-term stability of near-IR
spectrospcopic variations in L/T transition dwarfs is not well understood, so it
is possible that the 2012 September observations simply occurred during an epoch
of variability for PSO~J307.6+07, while the 2013 April observations found the
object in an epoch of constancy (or diminished variability).  It is also
possible that the detected variability is a result of an unrecognized oversight
in the observations or data reduction.  We note that the April data were
obtained at higher airmass; care was taken during these observations to ensure
the instrument slit was aligned with the parallactic angle, and the standard
stars were observed at similar airmasses (within 0.1).  We note as well that the
September 26th spectrum is the only one obtained in SXD (cross-dispersed) mode.
We have stitched the different orders together using the standard Spextool
procedure {\it xcombspec} to create a single spectrum, and it is possible that
this procedure has scaled one or more orders to an incorrect flux.  We consider
this to be unlikely for three reasons: \citet{Cushing:2005ed} showed that order
stiching with {\it xcombspec} results in a mean color difference of
$\langle\delta_{J-H}\rangle=0.00\pm0.04$ mag, much smaller than the $\Delta J$
we detected in PSO~J307.6+07; we used the same stitching procedure to obtain the
spectrum for PSO~J339.0+51 (Figure~\ref{fig.12-014644}), in which we find no
significant variation; and we see similar spectral variation in PSO~J140.2+45,
where both spectra were obtained in prism mode.  Overall, we cannot establish a
decisive reason for the difference of the intra-night behavior between our 2012
Sept and 2013 April datasets, so we label PSO~J307.6+07 a candidate
spectroscopic variable.  Further observations on multiple nights are needed to
confirm or reject this object as a near-IR variable.

\subsection{Candidate Photometric Variables}
\label{photvar} Since our seven discoveries are bright, they are all well
detected in 2MASS.  We can compare our MKO photometry to the 2MASS measurements
as a simple two-epoch check for near-IR variability over a period of
approximately ten years.  (Note that we do not use our SpeX data to synthesize
the conversions directly for each object, because the spectra and photometry
were all obtained at different epochs.)  We used the polynomial relations of
\citet[][Table 4]{Stephens:2004br} to bring the two epochs of $JHK$ photometry
onto a common system.  The UKIRT/WFCAM $JHK$ filters were designed on the MKO
photometric system, so we converted the 2MASS photometry for our discoveries to
MKO.  We find that the 2MASS and MKO $JHK$ magnitudes of our objects are
consistent within 0.1~mag, with two exceptions described below.

\subsubsection{PSO~J272.4689$-$04.8036 (WISE~180952.53$-$044812.5)}
\label{272discuss}
We assign a spectral type of T1 to PSO~J272.4689$-$04.8036 (hereinafter PSO
272.4$-$04), although it is fainter in the $H$-band than the T1 standard
SDSS~J0151+1244 \citep{Geballe:2002kw,Burgasser:2006cf}.  M13 classify this
object as a T0.5 (WISE~1809$-$0448).  Its $J_{\rm MKO}$ magnitude, as converted
from its 2MASS photometry, is $14.94\pm0.06$~mag, about 0.2~mag brighter than
our UKIRT measurement of $J_{\rm MKO}=15.15\pm0.01$~mag, a sign of potential
$J$-band variability.  The $K_{\rm MKO}=13.99\pm0.06$~mag (converted from 2MASS)
and VISTA $K_{\rm MKO}=13.98\pm0.01$~mag measurements are consistent.

\subsubsection{PSO~J339.0734+51.0978 (WISE~223617.59+510551.9)}
\label{339discuss}
We find PSO~J339.0734+51.0978 (hereinafter PSO~J339.0+51) to be a good match to
the T5 standard 2MASS~J1503+2525 \citep{Burgasser:2003ij,Burgasser:2006cf}.
Analysis of spectral indices gives a mean spectral type of T$5.1\pm0.1$, so we
assign a spectral type of T5, a half-type earlier than the T5.5 (WISE~2236+5105)
assigned by M13.  The $J_{\rm MKO}$ magnitude, converted from 2MASS, is
$14.31\pm0.04$~mag, about 0.15~mag brighter than the UKIRT $J_{\rm
  MKO}=14.46\pm0.01$~mag, so this object is another potential $J$-band variable.
The $H$-band magnitudes of $H_{\rm MKO}=14.53\pm0.05$~mag (converted from 2MASS)
and UKIRT $H_{\rm MKO}=14.62\pm0.02$~mag are consistent within $2\sigma$. We do
not have $K_{\rm MKO}$ photometry for PSO~J339.0+51.  We note that
  this object does not show any difference in spectra taken ten nights apart
  (Figure~\ref{fig.12-014644}).

\subsection{Candidate Binaries}
\label{binaries}
\begin{figure}
  \centering
  \includegraphics[width=0.6\linewidth, trim = 0 -2mm 0 0]{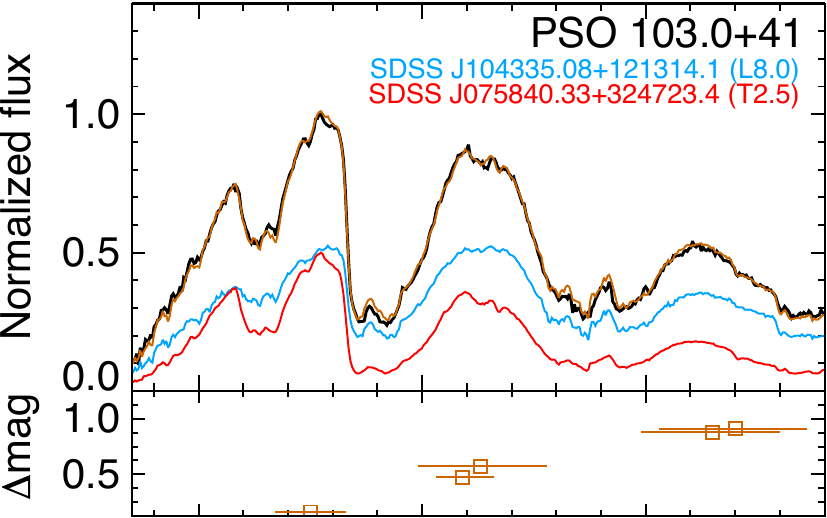}
  \includegraphics[width=0.6\linewidth]{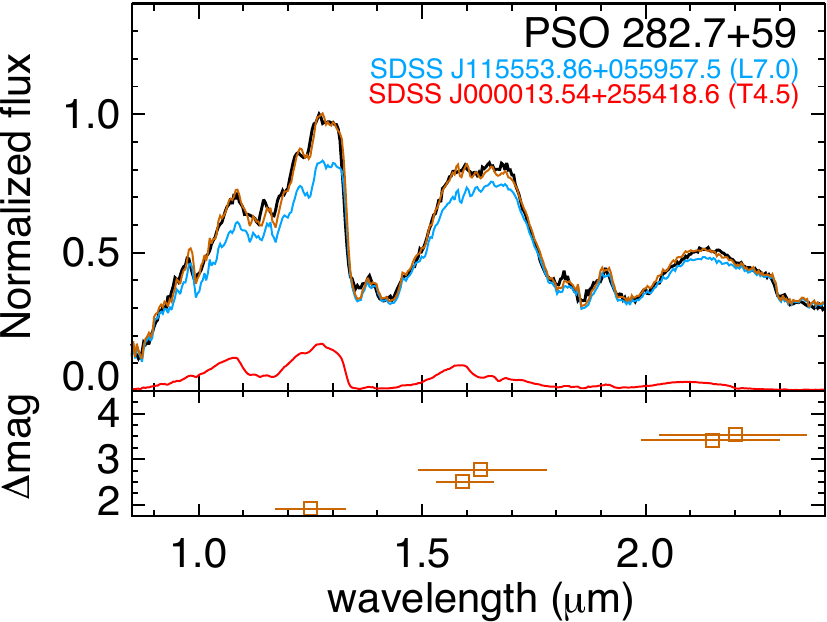}
  \caption{Best-matching template pairings for our candidate binaries PSO
    J103.0+41 ({\it top}) and PSO J282.7+59 ({\it bottom}).  Observed spectra
    are shown in black, individual component templates in blue and red, and
    convolved templates in brown.  The lower subpanels show the resulting flux
    ratios over standard NIR bandpasses computed from the best-matching template
    pairs (open brown squares).}
\label{fig.decomp}
\end{figure}

We have examined our seven discoveries for unusual spectral features that might
suggest unresolved binarity, by comparing their spectral indices to the
index-index and index-spectral type plots in \citet[][hereinafter
B10]{Burgasser:2010df}.  We found evidence for binarity in two of our
discoveries, described individually below.  We performed spectral decomposition
analysis on these two objects following the method described in Section~5.2 of
\citet{Dupuy:2012bp}.  Briefly, we used the library of 178 IRTF/SpeX prism
spectra presented in B10 to create summed spectra.  For each template pairing we
determined the scale factors needed to minimize the $\chi^2$ of the difference
with our observed spectrum.  We then examined the resulting best pairing to
determine the component spectral types, taking into account factors such as
larger than average spectral type uncertainties in the best-match templates and
the full range of properties implied when there were multiple matches giving
equally good fits.  We estimated the flux ratios in standard near-infrared
bandpasses using our $\chi^2$ values and the weighting scheme described in
\citet{Burgasser:2010df}.  Our derived component spectral types and their
corresponding uncertainties are listed in Table~\ref{tbl.decomp}, and the best
template pairing for each binary is shown in Figure~\ref{fig.decomp}.  (Our
analysis does not assess whether binary templates are better matches to our
observed spectra than single-object templates.)  For these template pairs, we
calculate photometric distances by convolving absolute magnitudes for each
component spectral type, determined using the same method as in
Section~\ref{results}.

\subsubsection{PSO~J103.0927+41.4601}
\label{103discuss}
Visually, the spectrum of PSO~J103.0927+41.4601 (hereinafter PSO~J103.0+41)
appears to be a slightly earlier spectral type than the T0 standard
SDSS~J1207+0244 \citep{Knapp:2004ji,Burgasser:2006cf}.  The spectral indices,
however, indicate a spectral type of T0.5 for PSO~J103.0+41.  Assuming this is a
single object, we settle on a classification of T0, placing the object at a
photometric distance of $14.2\pm1.2$~pc.  However, the closest spectral match is
in fact the L6+T2 binary SDSS~0423$-$0414AB
\citep{Geballe:2002kw,Burgasser:2005gj,Dupuy:2012bp}, and we find excellent
agreement with a template pairing of L8+T2.5 (Figure~\ref{fig.decomp},
Table~\ref{tbl.decomp}), which would indicate a distance of $20.1\pm2.4$~pc.
PSO~J103.0+41 also lies on the border of the binary regions in three of the B10
index-index plots.  This object is therefore an appealing candidate for
high-resolution imaging.

\subsubsection{PSO~J282.7576+59.5858}
\label{282discuss}
PSO~J282.7576+59.5858 (hereinafter PSO~J282.7+59) is not a good match to any of
the T~dwarf spectral standards. Its spectral indices average to L8.5, and our
best visual $J$-band fit is L9, but the $H$- and $K$-band peaks have lower flux
than the L9 standard 2MASS~J0255$-$4700
\citep{Martin:1999er,Kirkpatrick:2010dc}.  The shape of the $H$-band peak is
almost flat with a slight increase toward longer wavelengths, a feature more
typical of early L-dwarfs \citep[see Figure 10 in][]{Kirkpatrick:2010dc}.  We
tentatively classify PSO~J282.7+59 as an L9 for a single object (photometric
distance $13.0\pm1.1$~pc), but the discordance in its spectral features suggests
that we are looking at a blended spectrum. The best spectral match among known
ultracool dwarfs is actually the suspected triple system DENIS-P~0205$-$1159
\citep{Bouy:2005de}.  The spectral indices of PSO~J282.7+59 do not place it in
any of the binary regions of the B10 index-index plots, but B10 noted that their
selection process was probably biased against systems with more than two
components.  We find the best match for binaries has a template pairing of
L7+T4.5 (Figure~\ref{fig.decomp}, Table~\ref{tbl.decomp}), at a distance of
$17.9\pm2.1$~pc.

\subsection{Our Discoveries in Other Surveys}
\label{others}
All seven of the newly discovered objects are brighter than most known L/T
transition dwarfs.  This raises the question of why the objects were not
discovered earlier.  One reason is that several of the major surveys used in
previous brown dwarf searches covered smaller fractions of the sky than PS1.
Of our seven objects, only two are in regions covered by SDSS
  (PSO~J103.0+41 and PSO~J142.2+45) and only two lie within the UKIDSS search
  area (PSO~J007.7921+57.8267 and PSO~J142.2+45 again).  Another reason is that
we have searched close to the Galactic plane, using proper motion as well as
colors to distinguish brown dwarf candidates from reddened background objects.
Three of our seven discoveries have $|b|<10^{\circ}$, putting them in crowded
regions commonly avoided in previous searches \citep[e.g.,][]{Burgasser:2004hg}.
Our combination of \yps\ and $WISE$ photometry has also enabled us to discover
objects whose $JHK$ colors were excluded in near-infrared searches
\citep[e.g.,][]{Cruz:2003fi}.

\section{Summary}
\label{summary}
We have discovered seven L/T transition dwarfs within $9-15$~pc of the Sun based
on a search of the combined \PS\ $3\pi$ and \WISE\ databases.  We highlight
several specific objects:
\begin{itemize}
\item PSO~J140.2+45 (L9.5) and PSO~J307.6+07 (T1.5) show changes in their
  $J$-band spectra taken on different nights.  PSO~J140.2+45 also shows a
  marginal change in $Y$-band flux.  If confirmed by subsequent
    observations, these would be the third and fourth known near-infrared L/T
    transition variables, and PSO~J307.6+07 would be the second brightest,
    following SIMP~0136+0933.  PSO~J307.6+07 also has unusually blue $y_{\rm
    P1}-W1$ and $y_{\rm P1}-J$ colors compared to previously known objects of
  similar spectral type.
\item PSO~J272.4$-$04 (T1) and PSO~J339.0+51 (T5) have inconsistent 2MASS and UKIRT $J$
  magnitudes (by 0.2 and 0.15~mag, respectively), possibly signs of $J$-band
  variability.
\item PSO~J103.0+41 (T0) and PSO~J282.7+59 (L9) show spectral evidence that they are not
  single objects, including similarity to previously known multiple systems.
  High resolution spatial and spectroscopic observations are needed to
  investigate the possible multiplicity of these objects.
\end{itemize}

These brown dwarfs are all relatively bright ($J<15.5$) and easily close enough
for accurate parallax measurements.  All seven are excellent targets for
observations studying the atmospheric processes endemic to the L/T transition.

We thank the referee for helpful comments that improved the quality of this
paper.  The \PS\ Surveys (PS1) have been made possible through contributions of
the Institute for Astronomy, the University of Hawaii, the Pan-STARRS Project
Office, the Max Planck Society and its participating institutes, the Max Planck
Institute for Astronomy, Heidelberg and the Max Planck Institute for
Extraterrestrial Physics, Garching, The Johns Hopkins University, the University
of Durham, the University of Edinburgh, Queen’s University Belfast, the
Harvard-Smithsonian Center for Astrophysics, the Las Cumbres Observatory Global
Telescope Network, Inc., the National Central University of Taiwan, the Space
Telescope Science Institute, the National Aeronautics and Space Administration
under Grant No. NNX08AR22G issued through the Planetary Science Division of the
NASA Science Mission Directorate, and the University of Maryland.  We thank the
PS1 observing and processing staff for obtaining the PS1 data, and Bill Sweeney,
Heather Flewelling, and Mark Huber for assistance with accessing PS1 images.
The United Kingdom Infrared Telescope is operated by the Joint Astronomy Centre
on behalf of the Science and Technology Facilities Council of the U.K. This
paper makes use of observations processed by the Cambridge Astronomy Survey Unit
(CASU) at the Institute of Astronomy, University of Cambridge. We thank Mike
Irwin, Simon Hodgkin, and the team at CASU for making the reduced WFCAM data
available promptly, and Watson Varricatt and the UKIRT staff for carrying out
our observations.  We also thank John Rayner and Alan Tokunaga for making
engineering time available for IRTF observations, and Katelyn Allers, Michael
Kotson, Brian Cabreira, Bill Golisch, Dave Griep, and Eric Volqardsen for
assisting with these.  This project makes use of data products from the
Wide-field Infrared Survey Explorer, which is a joint project of the University
of California, Los Angeles, and the Jet Propulsion Laboratory/California
Institute of Technology, funded by the National Aeronautics and Space
Administration.  This research has made use of the 2MASS data products; the
UKIDSS data products; the VISTA data products; NASA's Astrophysical Data System;
the SIMBAD database operated at CDS, Strasbourg, France, the SpeX Prism Spectral
Libraries, maintained by Adam Burgasser at
http://pono.ucsd.edu/~adam/browndwarfs/spexprism, and the Database of Ultracool
Parallaxes, maintained by Trent Dupuy at https://www.cfa.harvard.edu/\mytilde
tdupuy/plx.  WMJB is supported by NSF grant AST09-09222.  Finally, the authors
wish to recognize and acknowledge the very significant cultural role and
reverence that the summit of Mauna Kea has always held within the indigenous
Hawaiian community. We are most fortunate to have the opportunity to conduct
observations from this mountain.

{\it Facilities:} \facility{IRTF (SpeX)}, \facility{PS1}, \facility{UKIRT (WFCAM)}

\newpage

\clearpage

\begin{deluxetable}{lccccccccc}
\tablecolumns{9}
\tablewidth{0pc}
\rotate
\tablecaption{IRTF/SpeX Observations \label{tbl1}}
\tablehead{   
  \colhead{Object} &
  \colhead{Date} &
  \colhead{Conditions} &
  \colhead{Seeing} &
  \colhead{Airmass} &
  \colhead{Mode} &
  \colhead{Slit} &
  \colhead{R} &
  \colhead{$T_\mathrm{int}$} & 
  \colhead{Standard} \\
  \colhead{} &
  \colhead{(UT)} &
  \colhead{} &
  \colhead{(arcsec)} &
  \colhead{} &
  \colhead{} &
  \colhead{(arcsec)} &
  \colhead{($\equiv \lambda/\Delta\lambda$)} &
  \colhead{(s)} &
  \colhead{} 
}
\startdata
PSO~J007.7921+57.8267 & 2012 Sep 24 & Clear & 0.9 & 1.27 & prism & $0.8\times15$ & $\sim$100 & 200 & HD 240290 \\
PSO~J103.0927+41.4601 & 2012 Sep 26 & Clear & 0.5 & 1.18 & prism & $0.5\times15$ & $\sim$100 & 960 & HD 39250 \\
PSO~J140.2308+45.6487 & 2012 Nov 8 & Clear & 0.8 & 1.32 & prism & $0.8\times15$ & $\sim$100 & 720 & HD 33654 \\ 
PSO~J140.2308+45.6487 & 2013 Jan 25 & Clear & 0.8 & 1.23 & prism & $0.8\times15$ & $\sim$100 & 180 & HD 79108 \\ 
PSO~J272.4689-04.8036 & 2012 Oct 14 & Clear & 1.1 & 1.28 & prism & $0.8\times15$ & $\sim$100 & 720 & HD 173591 \\
PSO~J282.7576+59.5858 & 2012 Sep 26 & Clear & 0.4 & 1.39 & prism & $0.8\times15$ & $\sim$100 & 240 & HD 240290 \\
PSO~J307.6784+07.8263 & 2012 Sep 20 & Clear & 0.6 & 1.15 & prism & $0.8\times15$ & $\sim$100 & 80 & HD 189920 \\
PSO~J307.6784+07.8263 & 2012 Sep 26 & Clear & 0.5 & 1.02 & SXD & $0.8\times15$ & $\sim$750 & 1440 & HD 189920 \\
PSO~J307.6784+07.8263 & 2013 Apr 03 & Clear & 1.0 & 1.43 & prism & $0.8\times15$ & $\sim$100 & 180 & HD 187170 \\
PSO~J307.6784+07.8263 & 2013 Apr 04 & Clear & 0.5 & 1.43 & prism & $0.8\times15$ & $\sim$100 & 480 & HD 187170 \\
PSO~J307.6784+07.8263 & 2013 Apr 05 & Cloudy & 0.8 & 1.50 & prism & $0.8\times15$ & $\sim$100 & 320 & HD 180150 \\
PSO~J339.0734+51.0978 & 2012 Oct 7 & Cloudy & 0.9 & 1.19 & prism & $0.8\times15$ & $\sim$100 & 1200 & HD 222749 \\
PSO~J339.0734+51.0978 & 2012 Oct 17 & Cloudy & 0.5 & 1.18 & SXD & $0.8\times15$ & $\sim$750 & 1920 & HD 209932 \\
\enddata
\end{deluxetable}

\newpage

\begin{deluxetable}{lcccc}
\tablecolumns{5}
\tablewidth{0pc}
\tabletypesize{\scriptsize}
\tablecaption{Properties of New Discoveries \label{tbl2}}
\tablehead{   
  \colhead{Property} &
  \colhead{PSO~J007.7921+57.8267} &
  \colhead{PSO~J103.0927+41.4601} &
  \colhead{PSO~J140.2308+45.6487\tablenotemark{a}} &
  \colhead{PSO~J272.4689$-$04.8036\tablenotemark{b}}
}
\startdata
\cutinhead{Astrometry}
PS1 R.A. (J2000) & $7.7921^\circ$ & $103.0927^\circ$ & $140.2308^\circ$ & $272.4689^\circ$ \\
 & ${\rm 00^h31^m10^s\!.11}$ & ${\rm 06^h52^m22^s\!.25}$ & ${\rm 09^h20^m55^s\!.40}$ & ${\rm 18^h09^m52^s\!.53}$ \\
PS1 Dec. (J2000) & $+57.8267^\circ$ & $+41.4601^\circ$ & $+45.6487^\circ$ & $-4.8036^\circ$ \\
 & $+57^\circ49'36.3''$ & $+41^\circ27'36.2''$ & $+45^\circ38'55.3''$ & $-04^\circ48'13.0''$ \\
PS1 epoch & 2011.410 & 2011.400 & 2011.483 & 2011.020 \\
Galactic longitude & $120.2261^\circ$ & $174.8199^\circ$ & $174.3678^\circ$ & $23.8612^\circ$ \\
Galactic latitude & $-4.9385^\circ$ & $17.7353^\circ$ & $44.7239^\circ$ & $6.9942^\circ$ \\
$\mu_\alpha$\,cos\,$\delta$ (mas\,yr$^{-1}$) & $523\pm17$ & $-8\pm6$ & $-42\pm23$ & $-62\pm18$ \\
$\mu_\delta$ (mas\,yr$^{-1}$) & $-1\pm16$ & $-38\pm6$ & $-843\pm23$ & $-429\pm17$ \\
2MASS designation & J00310928+5749364 & J06522224+4127366  & J09205549+4539058 & J18095256-0448081 \\
2MASS epoch & 1998.983 & 1998.268 & 1999.139 & 1998.747 \\
\WISE\ designation & J003110.04+574936.3 & J065222.24+412736.1 & J092055.40+453856.3 & J180952.53-044812.5 \\
\WISE\ epoch & 2010.316 & 2010.233 & 2010.307 & 2010.224 \\
\cutinhead{Photometry}
PS1 $z$ (AB mag) & $18.24\pm0.01$ & $18.85\pm0.02$ & $18.50\pm0.01$ & $18.80\pm0.03$ \\
PS1 $y$ (AB mag) & $17.01\pm0.01$ & $17.64\pm0.01$ & $17.21\pm0.01$ & $17.46\pm0.02$ \\
2MASS $J$ (mag) & $14.95\pm0.04$ & $15.48\pm0.06$ & $15.22\pm0.05$ & $15.14\pm0.06$ \\
2MASS $H$ (mag) & $13.78\pm0.04$ & $14.46\pm0.05$ & $14.16\pm0.05$ & $14.28\pm0.05$ \\
2MASS $K_s$ (mag) & $13.22\pm0.03$ & $13.89\pm0.05$ & $13.73\pm0.05$ & $13.96\pm0.06$ \\
MKO $Y$ (mag) & \nodata & \nodata & \nodata & \nodata \\
MKO $J$ (mag) & $14.79\pm0.01$\tablenotemark{c} & $15.36\pm0.01$\tablenotemark{c} & $15.04\pm0.01$\tablenotemark{c} & $15.15\pm0.01$\tablenotemark{d} \\
MKO $H$ (mag) & \nodata & $14.51\pm0.03$\tablenotemark{c} & $14.19\pm0.02$\tablenotemark{c} & \nodata \\
MKO $K$ (mag) & \nodata & $13.95\pm0.03$\tablenotemark{c} & $13.77\pm0.01$\tablenotemark{c} & $13.98\pm0.01$\tablenotemark{d,e} \\
\WISE\ $W1$ (mag) & $12.41\pm0.02$ & $13.13\pm0.02$ & $13.06\pm0.02$ & $13.29\pm0.03$ \\
\WISE\ $W2$ (mag) & $11.84\pm0.02$ & $12.44\pm0.03$ & $12.39\pm0.03$ & $12.73\pm0.03$ \\
\WISE\ $W3$ (mag) & $11.30\pm0.10$ & $11.80\pm0.25$ & $11.28\pm0.17$ & $12.38\pm0.47$ \\
\cutinhead{Spectral Indices}
H$_2$O-$J$ & $0.659$ (L8.5) & $0.588$ (T0.4) & $0.647$ (L8.9) & $0.657$ (L8.6) \\
CH$_4$-$J$ & $0.835$ ($<$T0) & $0.709$ ($<$T0) & $0.875$ ($<$T0) & $0.773$ ($<$T0) \\
H$_2$O-$H$ & $0.665$ (L7.8) & $0.575$ (T0.6) & $0.608$ (L9.7) & $0.630$ (L9.0) \\
CH$_4$-$H$ & $1.047$ ($<$T1) & $0.981$ (T1.1) & $1.002$ ($<$T1) & $0.972$ (T1.1) \\
CH$_4$-$K$ & $0.844$ (L8.7) & $0.778$ (L9.9) & $0.749$ (T0.3) & $0.719$ (T0.7) \\
H$_2$O-$K$ & $0.687$ (------) & $0.658$ (------) & $0.631$ (------) & $0.644$ (------) \\
\cutinhead{Physical Properties}
Near-IR spectral type & L9 & T0 & L9.5 & T1 \\
Photometric distance (pc) & $11.6\pm1.0$ & $14.2\pm1.2$ & $14.3\pm1.2$ & $15.0\pm1.3$ \\
$v_\mathrm{tan}$ (km\,s$^{-1}$) & $29\pm3$ & $3\pm1$ & $57\pm5$ & $31\pm3$ \\
Distinctive feature & ------ & Possible binary & $J$-band variable & Possible variable
\enddata
\tablenotetext{a}{First identified photometrically by \citet{Aberasturi:2011bk};
  also identified and spectrally typed by \citet{Mace:2013jh}.}
\tablenotetext{b}{Discovered independently by \citet{Mace:2013jh}.}
\tablenotetext{c}{Photometry obtained with UKIRT/WFCAM on 2012 December 13--14
UT.}
\tablenotetext{d}{Photometry obtained from the VISTA VHS catalog \citep{Cross:2012jz}.}
\tablenotetext{e}{VISTA uses a $K_s$ filter similar to 2MASS.}
\end{deluxetable}

\begin{deluxetable}{lccc}
\tablecolumns{4}
\tablewidth{0pc}
\tabletypesize{\scriptsize}
\tablecaption{Properties of New Discoveries \label{tbl3}}
\tablehead{   
  \colhead{Property} &
  \colhead{PSO~J282.7576+59.5858} &
  \colhead{PSO~J307.6784+07.8263\tablenotemark{a}} &
  \colhead{PSO~J339.0734+51.0978\tablenotemark{a}}
}
\startdata
\cutinhead{Astrometry}
PS1 R.A. (J2000) & $282.7576^\circ$ & $307.6784^\circ$ & $339.0734^\circ$ \\
 & ${\rm 18^h51^m01^s\!.84}$ & ${\rm 20^h30^m42^s\!.81}$ & ${\rm 22^h36^m17^s\!.63}$ \\
PS1 Dec. (J2000) & $+59.5858^\circ$ & $+7.8263^\circ$ & $+51.0978^\circ$ \\
 & $+59^\circ35'08.9''$ & $+07^\circ49'34.6''$ & $+51^\circ05'52.0''$ \\
PS1 epoch & 2011.177 & 2010.757 & 2010.688 \\
Galactic longitude & $89.5093^\circ$ & $51.9998^\circ$ & $102.3625^\circ$ \\
Galactic latitude & $23.2059^\circ$ & $-17.9389^\circ$ & $-6.2959^\circ$ \\
$\mu_\alpha$\,cos\,$\delta$ (mas\,yr$^{-1}$) & $23\pm19$ & $659\pm8$ & $736\pm14$ \\
$\mu_\delta$ (mas\,yr$^{-1}$) & $412\pm19$ & $-113\pm9$ & $330\pm8$ \\
2MASS designation & J18510178+5935040  & J20304235+0749358 & J22361685+5105487 \\
2MASS epoch & 1999.432 & 2000.442 & 2000.760 \\
\WISE\ designation & J185101.83+593508.6 & J203042.79+074934.7 & J223617.59+510551.9 \\
\WISE\ epoch & 2010.344 & 2010.333 & 2010.498 \\
\cutinhead{Photometry}
PS1 $z$ (AB mag) & $18.35\pm0.01$ & $17.97\pm0.01$ & $19.13\pm0.05$ \\
PS1 $y$ (AB mag) & $17.15\pm0.02$ & $16.44\pm0.01$ & $17.25\pm0.01$ \\
2MASS $J$ (mag) & $14.94\pm0.04$ & $14.23\pm0.03$ & $14.58\pm0.04$ \\
2MASS $H$ (mag) & $13.97\pm0.04$ & $13.44\pm0.03$ & $14.49\pm0.05$ \\
2MASS $K_s$ (mag) & $13.46\pm0.05$ & $13.32\pm0.04$ & $14.45\pm0.09$ \\
MKO $Y$ (mag) & \nodata & $15.22\pm0.01$\tablenotemark{b} & $15.66\pm0.01$\tablenotemark{b} \\
MKO $J$ (mag) & $14.85\pm0.01$\tablenotemark{b} & $14.05\pm0.01$\tablenotemark{b} & $14.46\pm0.01$\tablenotemark{b} \\
MKO $H$ (mag) & $14.03\pm0.02$\tablenotemark{b} & $13.48\pm0.01$\tablenotemark{b} & $14.62\pm0.02$\tablenotemark{b} \\
MKO $K$ (mag) & \nodata & \nodata & \nodata \\
\WISE\ $W1$ (mag) & $12.65\pm0.02$ & $12.96\pm0.03$ & $13.84\pm0.03$ \\
\WISE\ $W2$ (mag) & $12.18\pm0.02$ & $12.12\pm0.03$ & $12.48\pm0.03$ \\
\WISE\ $W3$ (mag) & $11.23\pm0.07$ & $10.96\pm0.11$ & $11.02\pm0.08$ \\
\cutinhead{Spectral Indices}
H$_2$O-$J$ & $0.678$ (L8.0) & $0.625$ (L9.4) & $0.225$ (T5.1) \\
CH$_4$-$J$ & $0.652$ (T0.1) & $0.698$ ($<$T0) & $0.405$ (T5.0) \\
H$_2$O-$H$ & $0.650$ (L8.3) & $0.586$ (T0.4) & $0.335$ (T5.1) \\
CH$_4$-$H$ & $1.016$ ($<$T1) & $0.878$ (T1.7) & $0.410$ (T5.2) \\
CH$_4$-$K$ & $0.897$ (L7.5) & $0.548$ (T2.2) & $0.203$ (T5.3) \\
H$_2$O-$K$ & $0.705$ (------) & $0.579$ (------) & $0.443$ (------) \\
\cutinhead{Physical Properties}
Near-IR spectral type & L9 & T1.5 & T5 \\
Photometric distance (pc) & $13.5\pm1.1$ & $10.9\pm0.9$ & $9.4\pm0.8$ \\
$v_\mathrm{tan}$ (km\,s$^{-1}$) & $26\pm3$ & $35\pm3$ & $36\pm3$ \\
Distinctive feature & Possible binary & $J$-band variable & Possible variable
\enddata
\tablenotetext{a}{Discovered independently by \citet{Mace:2013jh}.}
\tablenotetext{b}{Photometry obtained with UKIRT/WFCAM on 2012 September 19--21
UT.}
\end{deluxetable}

\begin{deluxetable}{lcc}
\tablecolumns{3}
\tablewidth{0pc}
\tabletypesize{\scriptsize}
\tablecaption{Properties of Candidate Binaries \label{tbl.decomp}}
\tablehead{   
  \colhead{Property} &
  \colhead{PSO~J103.0927+41.4601} &
  \colhead{PSO~J282.7576+59.5858}
}
\startdata
Primary Spectral Type & L8$\pm1$ & L7$\pm1$ \\
Secondary Spectral Type & T2.5$\pm0.5$ & T4.5$\pm1$ \\
$\Delta J$ (2MASS mag) & $0.24\pm0.05$ & $2.53\pm0.78$ \\
$\Delta H$ (2MASS mag) & $0.57\pm0.05$ & $3.04\pm1.03$ \\
$\Delta K$ (2MASS mag) & $0.91\pm0.08$ & $2.31\pm0.65$ \\
$\Delta J$ (MKO mag) & $0.18\pm0.05$ & $1.81\pm0.37$ \\
$\Delta H$ (MKO mag) & $0.58\pm0.04$ & $2.53\pm0.77$ \\
$\Delta$CH$_4$s (MKO mag) & $0.48\pm0.04$ & $3.13\pm1.08$ \\
$\Delta K$ (MKO mag) & $0.95\pm0.09$ & $1.90\pm0.41$ \\
\enddata
\tablecomments{These spectral types and flux ratios are estimated from spectral
  decomposition.  They have not been directly measured.}
\end{deluxetable}

\end{document}